\documentclass[13pt,usenatbib]{mn2e}
\usepackage{graphicx,times}
\usepackage{bm}
\usepackage{epsf}
\usepackage{amsmath}
\usepackage{amssymb}
\usepackage{color}

\date{\today,~ $ $Revision: 0.9 $ $}
\def\la{\langle}
\def\ra{\rangle}
\def\n{\noindent}
\def\be{\begin{equation}}
\def\ee{\end{equation}}
\def\ben{\begin{eqnarray}}
\def\een{\end{eqnarray}}
\def\nn{\nonumber}
\def\oh{\hat\Omega}
\def\myC{{\cal C}}
\def\myf{\Psi}

\def\cb{\textcolor{black}}
\def\ah{\textcolor{black}}

\begin{document}

\onecolumn
\title[New Approaches to Probing Minkowski Functionals]
{New Approaches to Probing Minkowski Functionals}
\author[Munshi et al.]
{D. Munshi$^{1,2}$, J. Smidt$^{3,4}$, A. Cooray$^{4}$, A. Renzi$^{5,6,7}$, A. Heavens$^{8}$, P. Coles$^{1,2}$\\
$^{1}$School of Physics and Astronomy, Cardiff University, Queen's
Buildings, 5 The Parade, Cardiff, CF24 3AA, UK,\\
$^{2}$ School of Mathematical and Physical Sciences, University of Sussex, Brighton BN1 9QH, U.K.\\
$^{3}$ Los Alamos National Laboratory, Theoretical Division, P.O. Box 1663, Mail Stop B283, Los Alamos, NM 87545, USA,\\
$^{4}$ Department of Physics and Astronomy, University of California, Irvine, CA 92697, \\
$^{5}$ INFN, Sezione di Padova, Via Marzolo 8, 35131 Padova, Italy,\\
$^{6}$ Dipartimento di Fisica e Astronomia ``G. Galilei",
Universit\'a degli Studi di Padova , Via Marzolo 8, 35131 Padova, Italy,\\
$^{7}$ SISSA, Via Bonomea 265, Trieste, I-34136, Italy, \\
$^{8}$ Imperial Centre for Inference and Cosmology, Department of Physics,  Imperial College, Blackett Laboratory, Prince Consort Road, London SW7 2AZ, U.K., \\}
\maketitle
\begin{abstract}
We generalize the concept of the ordinary skew-spectrum to probe the effect of non-Gaussianity
on the morphology of Cosmic Microwave Background (CMB) maps in several domains: 
\ah{in real-space (where they are commonly known as {\em cumulant-correlators}), and in harmonic and needlet bases.  The essential aim is to retain more information than normally contained in these statistics, in order to assist in determining the source of any measured non-Gaussianity, in the same spirit as Munshi \& Heavens' (2010) skew-spectra were used to identify foreground contaminants to the CMB bispectrum in Planck data.} Using a perturbative series to
construct the Minkowski Functionals (MFs), we provide a pseudo-$\myC_\ell$ based approach in both harmonic and needlet representations
to estimate these spectra in the presence of a mask and inhomogeneous noise.
Assuming homogeneous noise we present approximate expressions for
error covariance for the purpose of joint estimation of these spectra. We present specific results
for four different models of primordial non-Gaussianity {\em local}, {\em equilateral}, {\em orthogonal} and {\em enfolded} models, as well
as non-Gaussianity caused by unsubtracted point sources. Closed form results of next-order corrections to MFs too are obtained
in terms of a quadruplet of kurt-spectra. We also use the method of {\em modal decomposition} of the bispectrum and trispectrum to reconstruct 
the MFs as an alternative method of reconstruction of morphological properties of CMB maps.
Finally, we introduce the odd-parity skew-spectra to probe the odd-parity bispectrum and
its impact on the morphology of the CMB sky. Although developed for the CMB, the generic results obtained
here can be useful in other areas of cosmology.
\end{abstract}
\section{Introduction}
\label{sec:intro}
The study of Cosmic Microwave Background radiation provides the cleanest window to probe the very early 
stages of the Universe's history. This can be used to probe the mechanism that generates seed
perturbations, which lead to the structure that we observe in the present-day Universe.
Recent observations by WMAP \footnote{http://map.gsfc.nasa.gov/} and Planck\footnote{http://www.rssd.esa.int/index.php?project=Planck}
\citep{planck} supports the basic predictions of inflationary scenarios.
\cb{Recent results from Planck favour adiabatic and almost
Gaussian seed perturbations.} Furthermore, in future the proposed Experimental Probe of Inflationary Cosmology (EPIC)
survey, or ESA's Cosmic Origin Explorer \ah{or Polarized Radiation Imaging and Spectroscopy Mission  (COrE, \cite{core},  PRISM\footnote{http://www.prism-mission.org/}}), fourth-generation CMB satellite mission concepts, are very important in furthering our knowledge of the Universe. 
See e.g. \cite{Kermish}
for the POLARBEAR\footnote{http://bolo.berkeley.edu/polarbear/} experiment. The
current generation of ground-based observations, namely the \cb{Atacama} Cosmology 
Telescope (ACT; see \cite{Niem} for ACTPol)\footnote{http://www.physics.princeton.edu/act/} as well as the South Pole Telescope (SPT; see \cite{Jmc} for SPTPol)\footnote {http://pole.uchicago.edu/} are already
providing important clues especially of the CMB 
{\em secondary} anisotropy at smaller angular scales i.e. below a few arc minutes.
\begin{figure}
\begin{center}
{\epsfxsize=8. cm \epsfysize=8.5 cm {\epsfbox[30 150 592 715]{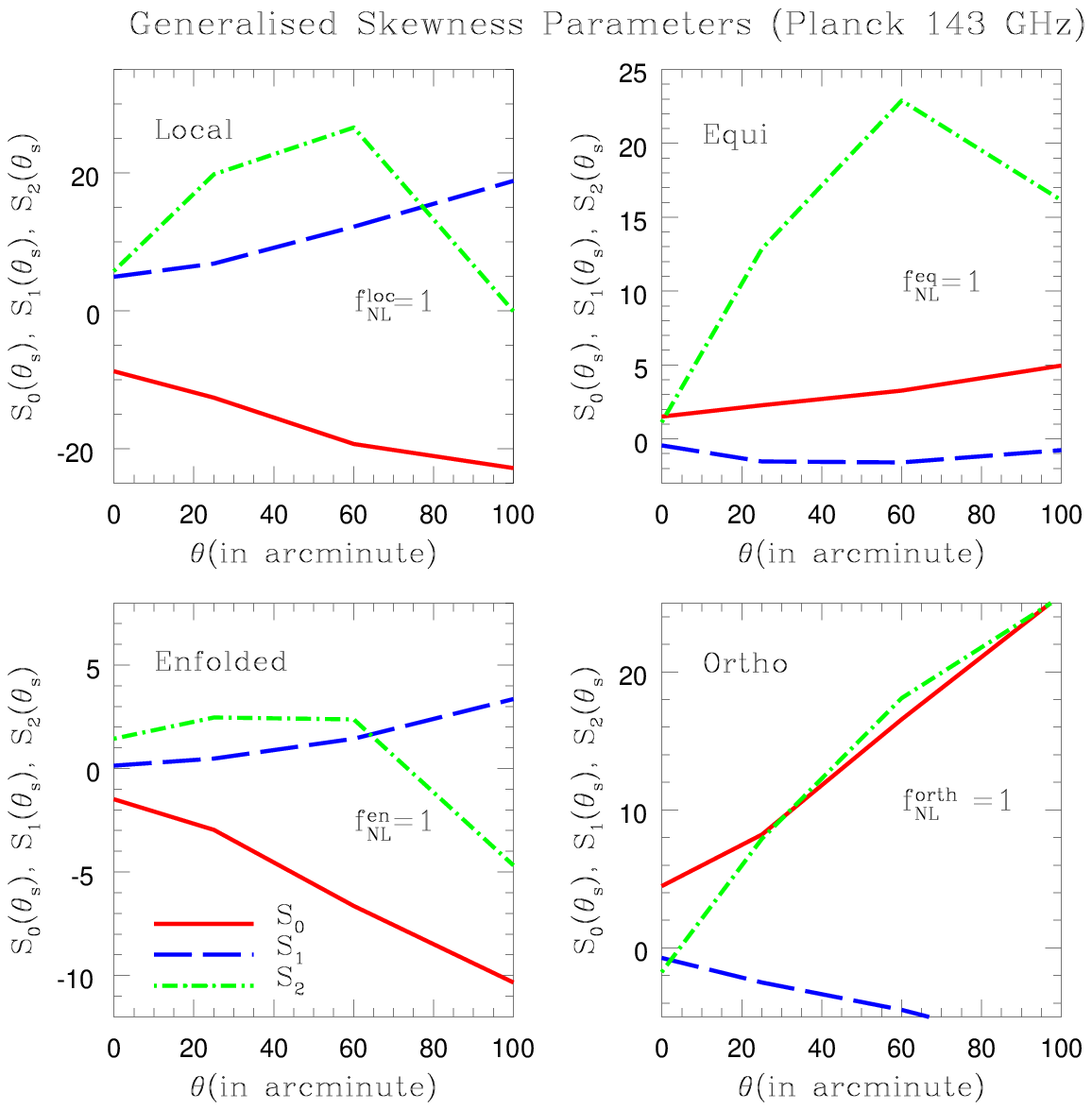}}} 
\caption{The generalised skewness parameters $S^{(0)}$ (solid-lines), $S^{(1)}$ (short-dashed lines)) and 
$S^{(2)}$ (long-dashed lines), defined in Eq. (\ref{skewness_real_space}), are plotted as a function of the beam FWHM $\theta_s$ (in arcminute).
The upper left panel corresponds to the {\em local} model and the upper
right panel correponds to the {\em equilateral} model, where as the lower-left and lower-right panel
correspond to {\em enfolded} and {\em orthogonal model} respectively.
The normalisation parameter  $f_{\rm NL}^{}$ describing various models
of non-Gaussianity are set to unity $f_{\rm NL}^{}=1$. The experimental setup corresponds to that of the Planck 143GHz channel.} 
\label{fig:skewness_onept}
\end{center}
\end{figure} 
It is well established now that non-Gaussianity from simplest inflationary models based on a 
single slowly-rolling scalar field is typically very small \citep{Salopek90,Salopek91,Falk93,Gangui94,Acq03,Mal03},  
(see e.g. \citet{Bartolo06} for a review). However, there are many variants of simple inflationary models 
which include models with multiple scalar fields \citep{lindemukha,Lyth03}, features in the inflationary potential, non-adiabatic fluctuations, non-standard kinetic terms, warm inflation \citep{GuBeHea02,Moss}, or
deviations from Bunch-Davies vacuum that can all lead to much higher level of non-Gaussianity, \ah{but these are heavily-constrained by Planck limits \cite{Planck13}}.

Early observational work on the bispectrum from COBE \citep{Komatsu02} and MAXIMA \citep{Santos} was followed by much more accurate analysis with WMAP \citep{Komatsu03,Crem07a,Spergel07} and Planck \citep{Planck13}.
Much of the interest in primordial non-Gaussianity has focused on a
phenomenological `{\em local} $f_{\rm NL}$' parametrization in terms of the perturbative non-linear
coupling in the primordial curvature perturbation \citep{verdeheavens}: 
\begin{equation}
\Phi(x) = \Phi_L(x) + f_{\rm NL} ( \Phi^2_L(x) - \langle \Phi^2_L(x) \rangle ) + g_{\rm NL}\Phi^3_L(x) 
+ h_{\rm NL}(\Phi^4_L(x)-3\la\Phi^2_L(x) \ra^2) + \cdots,
\label{eq:local}
\end{equation} 
where $\Phi_L(x)$ denotes the linear Gaussian part of the Bardeen curvature and $f_{\rm NL}, g_{\rm NL}, h_{\rm NL}$ are the
non-linear coupling parameters. A number of models have non-Gaussianity which can be
approximated by this form. The leading-order non-Gaussianity therefore is normally at the level of the
bispectrum, or in configuration space at the three-point level.
Many studies involving primordial non-Gaussianity have used the bispectrum, motivated by the
fact that it contains all the information about $f_{\rm NL}$ (\cite{Babich} but see \cite{KSH11}). It has been extensively
studied \cb{\citep{KSW,Crem03,Crem06,MedeirosContaldi06,Cabella06,Liguori07,YKW,Yadav08,SmSeZa09}}, with most of these measurements providing convolved estimates of the bispectrum. Optimized 3-point
estimators were introduced by \citet{Heav98}, and have been successively developed
\citep{KSW,Crem06,Crem07b,SmZaDo00,SmZa06} to the point where an estimator for $f_{\rm NL}$ which saturates
the Cramer-Rao bound exists for partial sky coverage and inhomogeneous noise \citep{SmSeZa09}.
Approximate forms also exist for {\em equilateral} non-Gaussianity, which may arise in models
with non-minimal Lagrangian with higher-derivative terms \citep{Chen06,Chen07}. In these models, the largest
signal comes from spherical harmonic modes with $\ell_1\simeq \ell_2 \simeq \ell_3$, whereas for
the local model, the signal is highest when one $\ell$ is much smaller than the other two --
the so-called {\em squeezed} configuration. The four different models that we consider
in this paper are {\em local}, {\em equilateral}, {\em orthogonal} and {\em enfolded} models of
primordial \cb{non-Gaussianity} (see e.g. \cite{MSC07, Komatsu_rev2010, YaWa10} \cb{for more detailed discussions about these models; a very short summary is provided in Appendix \ref{sec:mink_sky})}. 
 
While the bispectrum or its higher-order analogues, multispectra,  are more commonly used 
in studying departure from Gaussianity \citep{Bartolo04}, alternative statistics such as Minkowski functionals (MFs)
too are routinely used for this purpose. MFs describe the morphological features of a fluctuating (random) fields 
\citep{Mecke94,SB97,SG98,WK98}. The MFs for a Gaussian field are well understood and closed-form expressions 
exist~\citep{Tom86}. 
The MFs have been used to detect non-Gaussianity using projected (2D) fields such as the
CMB \citep{HKM06}, weak lensing \citep{MJ01} and 3D density fields as mapped by the galaxy surveys e.g. SDSS \citep{Pk05,Hk08,HKM06, HTS03, Hk02}. Several analytical results exist for prescriptions to model non-linear gravity as well as biasing schemes 
both in the quasilinear and highly nonlinear regimes \citep{Hk08}.
The MFs have been used also for the study of CMB data e.g. \cb{4-year COBE DMR data \citep{Nov00}}, BOOMERanG data \citep{Nat10} as well as for the WMAP data 
\citep{Komatsu03,Erik04,HM12}. The MF-based approach has also been used to study the effect of lensing
on the CMB \citep{STF00}.  
Using a MF-based approach on WMAP 7-year data \cite{HM12} recently obtained $f_{\rm NL}^{\rm loc}=20\pm 42$, $f_{\rm NL}^{\rm equi}=-121\pm 208$ and 
$f_{\rm NL}^{\rm orth}=-129\pm 171$.  
The recent constraints from Planck data release \citep{Planck13} are as follows: $f^{\rm NL}_{\rm local}=2.7 \pm 5.8, f^{\rm NL}_{\rm equil}= -42 \pm 75,$ and $f^{\rm NL}_{\rm ortho}= -25 \pm 39$. \cb{MF-based analysis using Planck data produces results
that are consistent with a null-hypothesis for the local model of non-Gaussianity \citep{Planck13a}.} 
While the estimation of primordial NG may be the primary motivation behind the study of MFs in
the context of CMB, they have also been applied to probe gravity-induced secondary NG
using weak-lensing convergence or $\kappa$-maps \citep{MJ01, Taruya20}. 
\begin{figure}
\begin{center}
{\epsfxsize=12. cm \epsfysize=6.5 cm {\epsfbox[30 423 590 715]{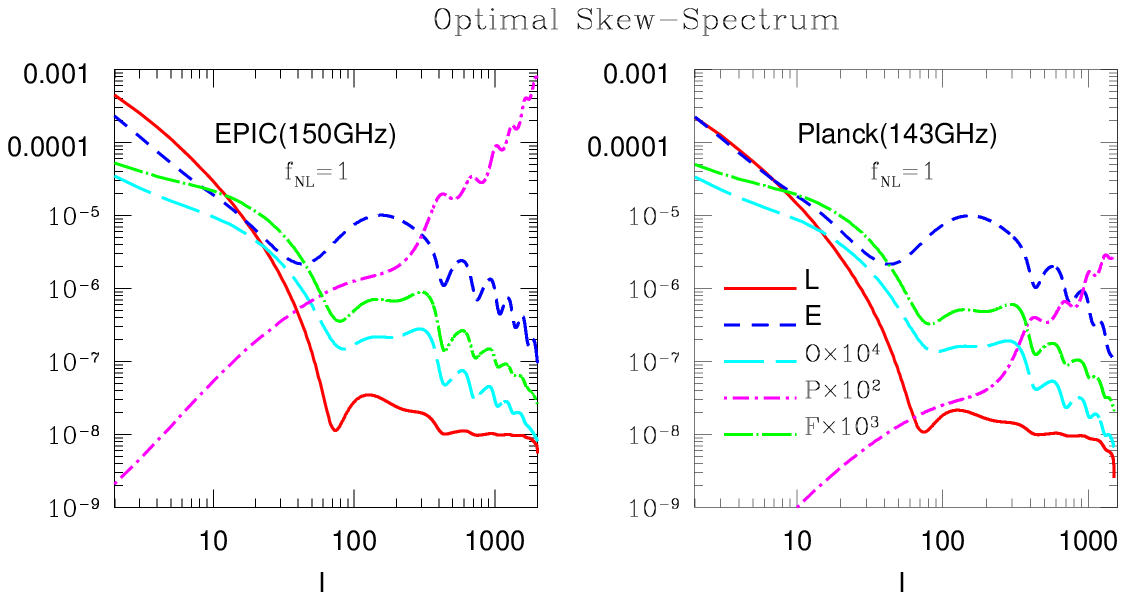}}}
\caption{\ah{The signal expected from the} optimum estimator (see Eq. (\ref{eq:muhe10}) for definition) for the skew-spectra ${\cal S}^{\rm opt}_{\ell}$ are shown as a function of the angular harmonics $\ell$. The left panel correspond to EPIC (150GHz channel) and the right panel correspond to Planck (143 GHz channel). 
The \cb{solid}, short-dashed, long-dashed, dot-short dashed and dot-long dashed lines correspond to
local (L), equilateral (E), orthogonal (O) and enfolded (F) and point-sources (P) respectively.
We have scaled some of the models as depicted. The parameter $f_{\rm NL}$ is set to unity for each of these models.}
\label{fig:opt}
\end{center}
\end{figure}
One of the main motivations behind studying various alternatives that probe primordial
non-Gaussianity has to do with issues related to estimation. Different probes are affected
differently by different contamination such as the presence of foreground or secondary
non-Gaussianity. The methods based on multispectra that are typically employed for the
estimation of non-Gaussianity use a Fourier (or harmonic) space approach.
On the other hand, the techniques developed for estimation of MFs are traditionally applied 
in real space. \cite{Mat03}
obtained a closed-form expression for MFs in $D$ dimensions using a perturbative 
expansion in terms of various orders of multispectra. At the lowest order, a 
departure from Gaussianity is characterized by three different skewness 
parameters $S^{(0)}, S^{(1)}, S^{(2)}$. The skewness parameter $S^{(0)}$ is 
the ordinary skewness that is most commonly used in various studies of non-Gaussianity
for projected surveys as well as in 3D. The other set of skewness parameters are
defined in terms of different cubic statistics that are constructed from the 
original data using differential operators.   \ah{The purpose of this paper is to retain more of the details of the non-Gaussianity in these methods, in order to 
aid in determining the source of non-Gaussianity.  This is very similar to the approach of \cite{MuHe10}, who devised an optimal compression of bispectrum data which retained enough information to determine point source and lensing-ISW contributions to Planck data \citep{Planck13}.}

This paper is organized as follows. In \textsection\ref{sec:form} we review the formalism of Minkowski Functionals (MFs). 
In \textsection\ref{sec:skew} we focus on the one-point estimators, the generalized skewness, and their links to Minkowski Functionals. In \textsection\ref{sec:estim}
real-life issues such as the mask and noise are discussed and estimators are designed for estimation of
the Minkowski Functionals using a pseudo-$\myC_{\ell}$ based approach. 
In \textsection\ref{sec:modal} we provide generic result to reconstruct the MFs using modal decomposition
of the bi- or trispectrum. In \textsection\ref{sec:odd} we extend the concept of skew-spectrum to odd-parity bispectra.
In \textsection\ref{sec:needlet} we present our results of morphological analysis in the needlet basis.
We relate the skew-spectrum defined in the needlet basis with that in harmonic domain. 
Finally, \textsection\ref{sec:disc} is left for discussion and \textsection\ref{sec:conclu} for conclusions.

In Appendix \ref{sec:mink_sky} various early Universe models and their predictions for the lower-order multispectra are presented. In  Appendix \ref{sec:kurt} the corrections to the lowest order in
non-Gaussianity are discussed. These corrections are related to estimation of trispectra. 

Throughout in this paper a WMAP7 \citep{WMAP7} background cosmology will be used for computation of various spectra.
Unless specified, the values of the normalisation  parameters $f_{\rm NL}^{\rm loc}$, $f_{\rm NL}^{\rm equi}$, $f_{\rm NL}^{\rm en}$ 
and $f_{\rm NL}^{\rm orth}$ (to be defined later), are set to unity.%
\footnote{The techniques presented in this paper have already been used in the context of CMB secondaries \citep{MuCoHe}, 
frequency-cleaned thermal Sunyaev Zeldovich (tSZ) map \citep{MuSmJouCo}, weak lensing convergence maps \citep{MuSmWaCo} as well as in galaxy redshift surveys   
\citep{PrMu}.}      

We will consider two different experimental setupts. For Planck we take the \cb{143}GHz channel and for EPIC
we take the 150GHz channel. The beams for these experiments are $\theta_s=7.1'$ and  $\theta_s=5'$.
The pixel areas for Planck and EPIC are given by $\Omega_{\rm pix}=0.0349$ and $\Omega_{\rm pix}=0.0002$.
The noise per pixel for Planck is $\sigma_{\rm pix}=2.2\times 10^{-6}$ and for EPIC is
\cb{given} by $\sigma_{\rm pix}=8.0\times 10^{-6}$.
\section{Formalism}
\label{sec:form}
The MFs are well-known morphological descriptors which are used in the study of random fields. 
Morphological properties are the properties that remain invariant under rotation and translation (see \cite{Hadwiger59}
for a more formal introduction). They are defined
over an excursion set $\Sigma$, for a given threshold $\nu$. The three MFs that are defined for two-dimensional (2D)
studies can be expressed as:
\be
{\cal V}_0(\nu) = \int_{\Sigma} da; \quad {\cal V}_1(\nu) = {1 \over 4}\int_{\partial\Sigma} dl; \quad {\cal V}_2(\nu) = {1 \over 2\pi}\int_{\partial \Sigma}\kappa dl.
\ee
\n
Here $da$, $dl$ are the elements for the excursion set $\Sigma$ and its boundary $\partial \Sigma$. The MFs 
${\cal V}_k(\nu)$
correspond to the area of the excursion set $\Sigma$, the length of its boundary $\partial\Sigma$, and the
integral curvature along its boundary, which is related to the genus $g$ and hence the Euler characteristic $\chi$.
\begin{figure}
\begin{center}
{\epsfxsize=13.5 cm \epsfysize=6. cm {\epsfbox[30 483 592 715]{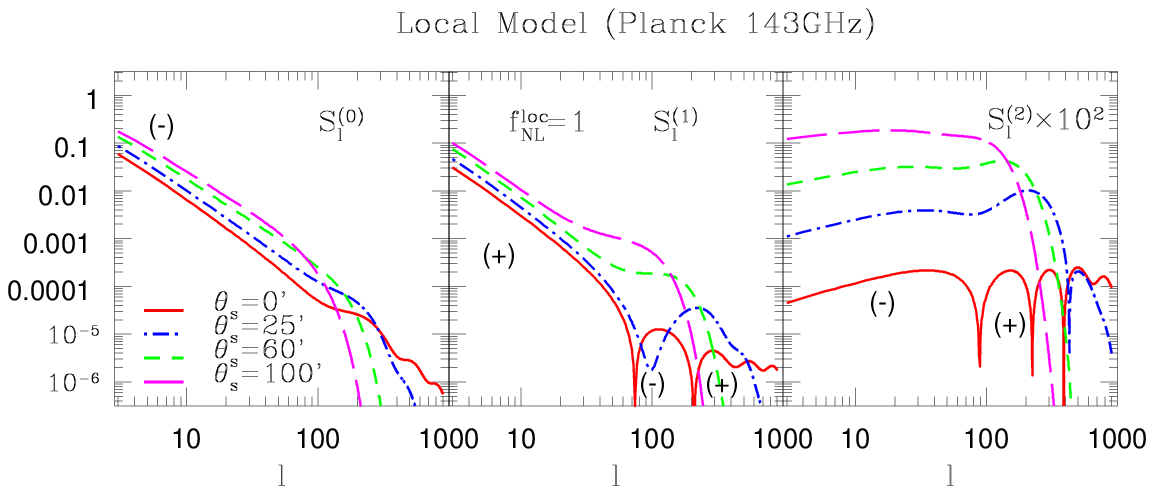}}}
\caption{The skew-spectra $S^{(0)}_\ell$ (left-panel), $S^{(1)}_\ell$(middle-panel) and $S^{(2)}_\ell$
(right-panel) are plotted for various smoothing \cb{beams} (see Eq. (\ref{eq:S_l})) \cb{as} function 
of the harmonics $\ell$. The model for the primordial non-Gaussianity is assumed to be a local model for these plots Eq. (\ref{eq:model_loc}).The underlying cosmology is assumed to
be that of WMAP7. The FWHM for the four beams considered are 
$\theta_s = 0'$ (solid), $25'$(dot-dashed), $50'$ (small-dashed), $100'$(long-dashed).
The parameter  $f_{\rm NL}^{\rm loc}$ describing the normalization of local model
of non-Gaussianity is set to unity $f_{\rm NL}^{\rm loc}=1$, and the resolution is set at $\ell_{\rm max}=1500$.
To extract the skewness parameters we use the relationship $S^{(i)}= {1 \over 4 \pi}\sum_\ell\Xi_{\ell}S_\ell^{(i)}$.
The experimental noise corresponds to that of Planck 143GHz channel.}
\label{fig:S_loc}
\end{center}
\end{figure}
In our analysis 
we will consider a smoothed random field  (e.g. CMB temperature distribution on the surface of the sky)
$\myf(\oh)={\delta T(\oh)/T_0}$ with mean $\la \myf(\oh)\ra=0$ and variance $\sigma_0^2 = \la \myf^2(\oh) \ra$,
for a generic 2D weakly non-Gaussian random field $\myf$. The spherical harmonic decomposition using
$Y_{\ell m}(\oh)$ as a basis function, $\myf(\oh) = \sum_{\ell m} \myf_{\ell m} Y_{\ell m}(\oh)$ can be used to define
the power spectrum $\myC_l$ which is sufficient characterization of a Gaussian field  
$\la \myf_{\ell m} \myf^*_{\ell' m'} \ra = \myC_\ell \delta_{\ell\ell'}\delta_{mm'}$.
For a non-Gaussian field the higher-order statistics such as the bi- or trispectrum can describe the resulting mode-mode coupling. 
Alternatively the topological measures such as Minkowski functionals which include the Euler characteristic or genus
can be employed to quantify deviation from Gaussianity. 
At the leading order the MFs can be constructed
completely from the knowledge of the bispectrum alone. We will be studying the MFs defined over the surface of the
celestial sphere, but equivalent results can be obtained in 3D using a Fourier decomposition (Munshi 2013, in preparation).
The notation and analytical results in this section are being kept generic, however they will be specialized to the case of the CMB sky in
subsequent discussions. \cb{Using a perturbative series expansion in the field r.m.s., $\sigma_0$, the MFs denoted as ${\cal V}_k(\nu)$ for a threshold $\nu= \myf/\sigma_0$
can be expressed as follows \citep{Mat03}}: 
\ben
&& \cb{ {\cal V}_k(\nu) = 
{1 \over (2\pi)^{(k+1)/2}} {\omega_2 \over \omega_{2-k}\omega_k} \exp \left ( -{\nu^2 \over 2}\right ) \left ( \sigma_1 \over \sqrt 2 \sigma_0 \right )^k
\left [{v}_k^{(1)}(\nu)+ {v}_k^{(2)}(\nu)\sigma_0 + {v}_k^{(3)}(\nu)\sigma_0^2 + {v}_k^{(4)}(\nu)\sigma_0^3 + \cdots \right ];} \label{eq:small_vs}\\
&& {v}_k^{(1)}(\nu) = {\cal H}_{k-1}(\nu); \quad  {v}_k^{(2)}(\nu) = \left [ \left \{  {1 \over 6} S^{(0)} {\cal H}_{k+2}(\nu) + {k \over 3} S^{(1)} {\cal H}_k(\nu) + {k(k-1) \over 6} S^{(2)} {\cal H}_{k-2}(\nu)\right \} \right ];  \label{eq:skewness_def}\\
&& \sigma_j^2 = {1 \over 4\pi }\sum_\ell \; {\Pi}_{\ell}^j\, \Xi_{\ell}\, \myC_\ell\, b_\ell^2; \quad 
\Pi_{\ell} = \ell(\ell+1); \quad \Xi_{\ell} = (2\ell+1).
\label{eq:v_k}
\een
\n
The constant $\omega_k$ introduced above is the volume of the unit sphere in $k$-dimensions: $w_k = {\pi^{k/2}\over \Gamma(k/2+1)}$, \ah{and the skewness parameters $S^{(i)}$ are defined below in Eq. (\ref{skewness_real_space}).}
In 2D we will only need $\omega_0=1$, $\omega_1=2$ and $\omega_2 =\pi$ and the lowest-order Hermite polynomials ${\cal H}_k(\nu)$
are listed below:
\ben
&& {\cal H}_{-1}(\nu) = \sqrt{\pi \over 2} \exp\left ({\nu^2 \over 2}\right ) {\rm erfc} \left (\nu \over \sqrt 2 \right ); \quad {\cal H}_0(\nu) = 1, \quad {\cal H}_1(\nu) = \nu, \nn \\
&& {\cal H}_2(\nu)=\nu^2 -1, \quad {\cal H}_3(\nu)=\nu^3 - 3\nu, \quad \quad {\cal H}_4(\nu) = \nu^4 - 6\nu^2 + 3; \nn \\
&& {\cal H}_n(\nu) = (-1)^n \exp \left ({ \nu^2 \over 2 } \right ) {\ah{d^n} \over d\nu^n} \exp \left (-{\nu^2 \over 2 }\right ).
\een

\cb{The MFs consist of two distinct contributions: one, which is independent of 
the three different skewness parameters $S^{(0)}, S^{(1)}, S^{(2)}$, signifies the MFs for a Gaussian random field, and are denoted by 
$v_k^{(1)}(\nu)$;
the other contribution ${v}^{(2)}_k(\nu)$ represents the departure from Gaussian statistics and depends on the
generalized skewness parameters.
The next-to-leading order corrections ${v}^{(3)}_k(\nu)$ depend on the generalised kurtosis parameters and will be discussed
in more detail in Appendix-{\ref{sec:kurt}}.} 
Various moments $\sigma_j$ that appear in Eq. (\ref{eq:v_k}) can be expressed in terms of the power spectrum $\myC_\ell$ and the observational
beam $b_\ell(\theta_s)$ (the full width at half maximum or FWHM is denoted by $\theta_s$). 
The moments that will mostly be used are $\sigma_0^2 = \la \myf^2 \ra $ and $\sigma_1^2 = \la (\nabla \myf)^2 \ra$.

The real-space expressions for the triplets of skewness $S^{(i)}$, which appear in the expressions for the MFs, are given below. These are natural generalizations of the ordinary skewness 
$S^{(0)}$ that is used in many cosmological studies, but are constructed from different cubic combinations.
\begin{figure}
\begin{center}
{\epsfxsize=13.5 cm \epsfysize=6.0 cm {\epsfbox[30 483 583 715]{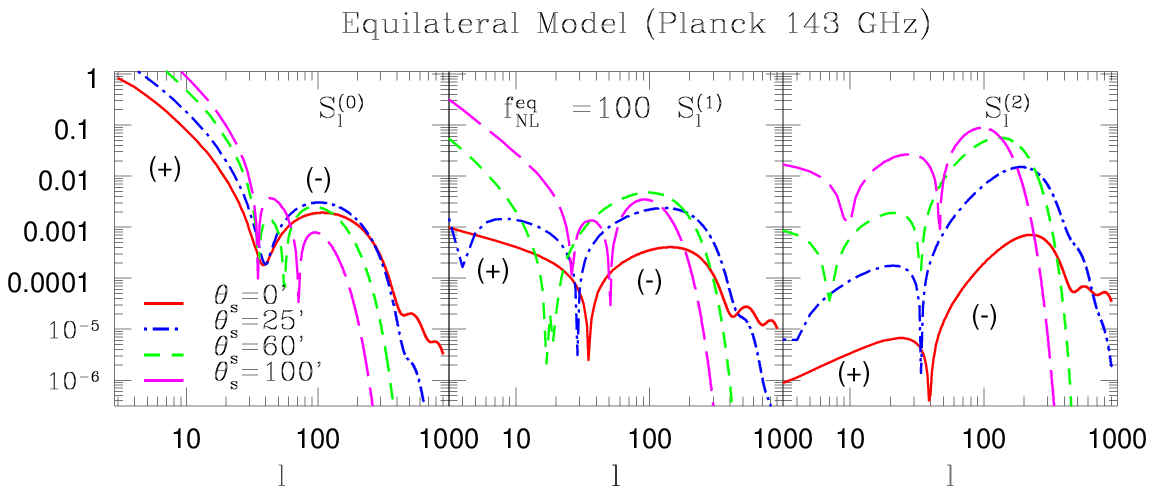}}}
\caption{Same as previous figure but for the equilateral model of primordial
non-Gaussianity as defined in Eq. (\ref{eq:model_eq}). The choice of smoothing angular scales (beam) are
same as previous figure. The normalization parameter for the equilateral bispectrum $f_{\rm NL}^{\rm eq}$ 
is fixed at $f_{\rm NL}^{\rm eq}=100$.} 
\label{fig:S_eq}
\end{center}
\end{figure}
\begin{figure}
\begin{center}
{\epsfxsize=13.5 cm \epsfysize=6.0 cm {\epsfbox[30 483 583 715]{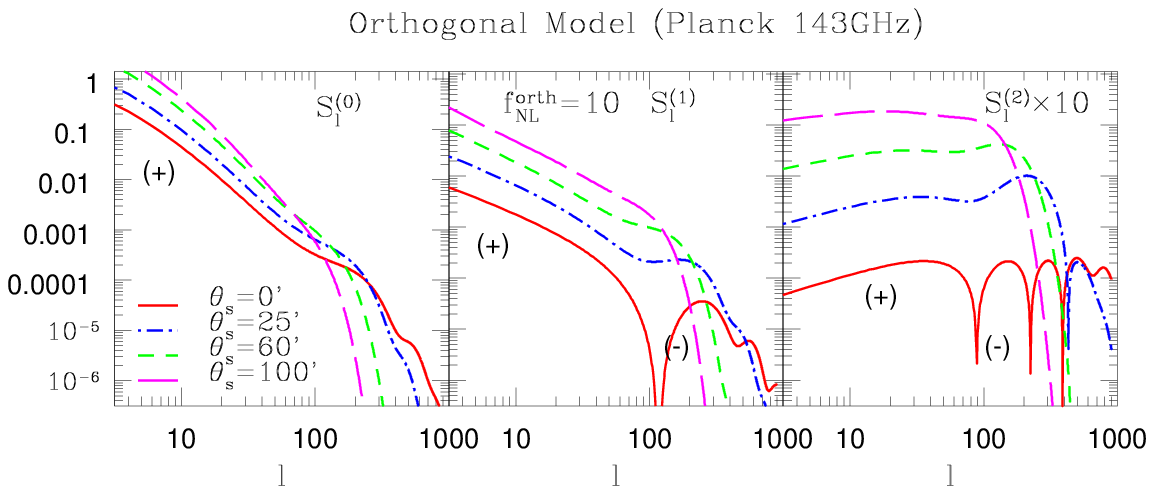}}}
\caption{Same as previous figure but for the orthogonal model of primordial
non-Gaussianity. The choice of smoothing angular scales (beam) are
same as previous figure. The value of the normalization parameter for the orthogonal model of bispectrum $f_{\rm NL}^{\rm orth}$ is fixed at
$f_{\rm NL}^{\rm orth}=10$.} 
\label{fig:S_or}
\end{center}
\end{figure}
\begin{figure}
\begin{center}
{\epsfxsize=13.5 cm \epsfysize=6.5 cm {\epsfbox[30 483 583 725]{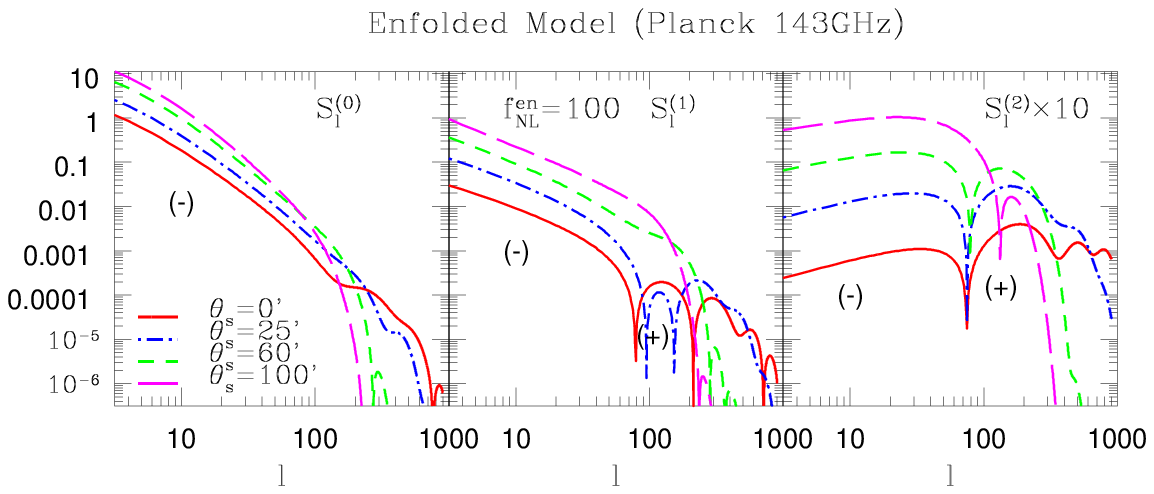}}}
\caption{Same as previous figure but for the enfolded model of primordial
non-Gaussianity. The choice of smoothing angular scales (beam) are
same as previous figure. The normalization parameter for the enfolded bispectrum $f_{\rm NL}^{\rm en}$ is fixed at
$f_{\rm NL}^{\rm en}=100$.} 
\label{fig:S_en}
\end{center}
\end{figure} 
\be
S^{(0)} \equiv {S^{(\myf^3)} \over \sigma_0^4} = {\la \myf^3 \ra \over \sigma_0^4}; \quad
S^{(1)} \equiv-{3 \over 4}{S^{(\myf^2\nabla^2 \myf)} \over \sigma_0^2\sigma_1^2} = -{3 \over 4}{\la \myf^2 \nabla^2 \myf \ra \over \sigma_0^2 \sigma_1^2}; \quad
S^{(2)} \equiv \cb{-3}{S^{(\nabla \myf \cdot \nabla \myf \nabla^2 \myf)}\over \sigma_1^4} = -{3}{\la (\nabla \myf).(\nabla \myf) (\nabla^2\myf) \ra  \over \sigma_1^4}.
\label{skewness_real_space}
\ee
These one-point generalised skewness parameters are plotted in Figure \ref{fig:skewness_onept} for various models
of non-Gaussianity. The expressions in the harmonic domain are more useful in the context of CMB studies where we will be recovering
them from a masked sky using analytical tools that are commonly used for power spectrum analysis. 
The expressions for the MFs in Eq. (\ref{eq:v_k}) that we have discussed, depend on the one-point cumulants $S^{(i)}$.  However, it is possible to define power spectra associated with each of these skewness parameters following a procedure similar to that 
developed in \cite{MuHe10}. 
This is one of the main motivations behind generalizing the concept of MFs, each of which is a number,
to a set of power spectra. As an illustration of the power of the skew-$\myC_{\ell}$s see \citep{Planck13}
which demonstrates detection of ISW-Lensing and point source non-Gaussianities.

The series expansion for the MFs can be extended beyond the level of the bispectrum.
The next-to-leading order corrections terms are related to trispectra of the original map.
These corrections are expected to be sub-dominant in the context of CMB studies for 
the entire range of angular scales being probed. \cb{However, if the primordial bispectrum
is negligible, as seems to be the case, these terms may play an important role in shaping the topology of the CMB sky. In addition,
lensing-induced topology change appears only at the level of trispectrum.}

The results here correspond to maps of temperature, which is a spin-$0$ object. It is possible to 
extend these results to the case of polarization analysis i.e. for spin-2 fields. Such results will
also be relevant in the context of weak lensing shear and flexions. A detailed analysis will be presented elsewhere. 
\begin{figure}
\begin{center}
{\epsfxsize=13.5 cm \epsfysize=6.0 cm {\epsfbox[10 483 583 715]{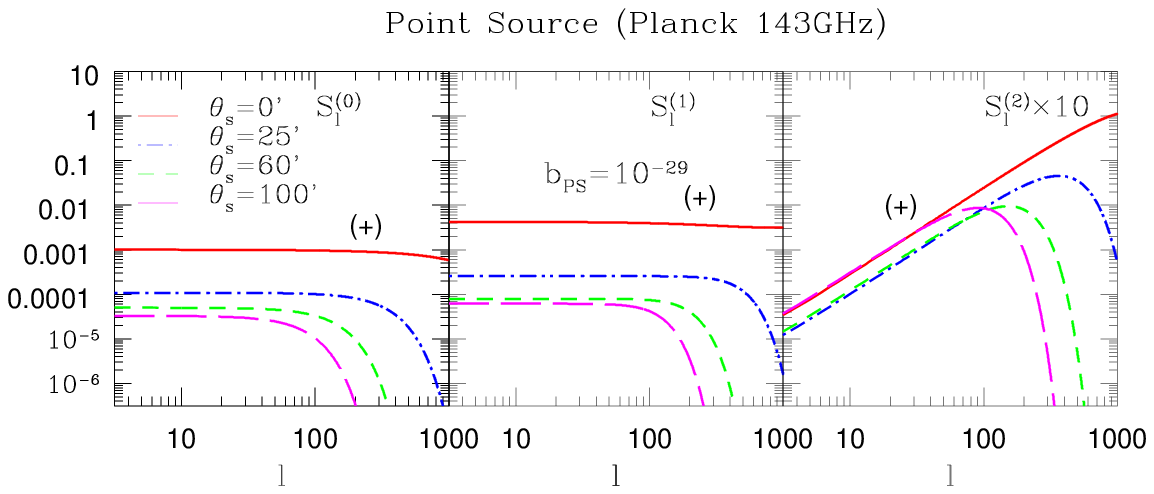}}}
\caption{Same as previous figures but for \ah{uncorrelated} point sources. For the point sources we have taken $b_{\rm PS}=10^{-29}\mu K^3$.
The panels represent the three different skew-spectra $S^{(0)}_\ell$ (left-panel), $S^{(1)}_\ell$ (middle panel) and 
$S^{(2)}_\ell$ (right-panel) respectively. The skew-spectra are sensitive to the resolution of the maps being analyzed
as they are integrated measures. This is related to the fact that it's value at a specific harmonics $\ell$ 
takes contribution from all possible modes, i.e. the entire range of $\ell$ values. The point source contributions
dominate the primordial non-Gaussianity at high $\ell$.}
\label{fig:S_ps}
\end{center}
\end{figure} 
\section{The triplets of Skew-Spectra and Lowest-Order Corrections to Gaussian MFs}
\label{sec:skew}
The skew-spectra are cubic statistics that are constructed by cross-correlating two different fields. 
One of the fields used is a composite field, typically a  product of two maps, either in its original form or constructed 
by means of relevant differential operations. The second field will typically 
be a single field but may be constructed by applying various differential operators. 
These three skewness parameters contribute to the three MFs that we will consider in 2D.

The first of the skew-spectra was studied by \cite{Cooray01}
and later by \cite{MuHe10} and is related to the commonly-used skewness.
The skewness in this case is constructed by cross-correlating the
squared map $\myf^2(\oh)$ with the original map $\myf(\oh)$.  The
second skew-spectrum is constructed by cross-correlating the squared map $\myf^2(\oh)$ with $\nabla^2\myf(\oh)$.
Analogously the third skew-spectrum represents the cross-spectra that can be constructed 
using $\nabla \myf(\oh)\cdot\nabla \myf(\oh)$ and $\nabla^2 \myf(\oh)$:
\ben
&& S_\ell^{(0)} \equiv {1 \over \sigma_0^4}{}S_\ell^{(\myf^2,\myf)} \equiv {1 \over \sigma_0^4}{1 \over \Xi_{\ell}}\sum_m 
{\rm Real}(\la [\myf]_{\ell m}[\myf^2]^*_{\ell m} \ra) 
={1 \over \sigma_0^4}{\sum_{\ell_1\ell_2} B_{\ell\ell_1\ell_2}J_{\ell\ell_1\ell_2}
b_{\ell}b_{\ell_1}b_{\ell_2}};  
\label{sl1} \\
&& S_\ell^{(1)} \equiv -{3 \over 4}{}{1 \over \sigma_0^2\sigma_1^2}S_\ell^{(\myf^2,\nabla^2\myf)} 
\equiv -{3 \over 4 \sigma_0^2\sigma_1^2}{1 \over \Xi_{\ell}}\sum_m
{\rm Real}(\la [\nabla^2 \myf]_{\ell m}[\myf^2]^*_{\ell m}\ra)  \nn \\
&& \quad\quad =-{3 \over 4 \sigma_0^2\sigma_1^2}{\sum_{\ell_1\ell_2}{1 \over 3}\Big ({\Pi}_{\ell} + {\Pi}_{\ell_1} + {\Pi}_{\ell_2} \Big )B_{\ell\ell_1\ell_2}J_{\ell\ell_1\ell_2}b_{\ell_1}b_{\ell_2}b_{\ell_3}}; \label{sl2}\\
&& 
S_\ell^{(2)} \equiv -{3 \over \sigma_1^4}{}S_\ell^{(\nabla \myf\cdot\nabla \myf, \nabla^2\myf)} \equiv 
-{3 \over \sigma_1^4}{1 \over \Xi_{\ell}}\sum_m
{\rm Real}(\la [\nabla \myf \cdot \nabla \myf]_{\ell m}[\nabla^2 \myf]^*_{\ell m} \ra) \nn \\
&& \quad\quad ={3 \over \sigma_1^4}{\sum_{\ell_1\ell_2}
{1 \over 3}\Big [ ({\Pi}_{\ell}+{\Pi}_{\ell_1} - {\Pi}_{\ell_2}){{\Pi}_{\ell_2} \over 2} + {\rm cyc.perm.} \Big ]
 B_{\ell\ell_1\ell_2}J_{\ell\ell_1\ell_2}b_{\ell}b_{\ell_1}b_{\ell_2}}; \label{sl3}\\
&& J_{\ell_1\ell_2\ell_3} \equiv {1 \over \Xi_{\ell_3}}I_{\ell_1\ell_2\ell_3} = \sqrt{\Xi_{\ell_1}\Xi_{\ell_2} \over 4\pi \; \Xi_{\ell_3} }\left ( \begin{array}{ c c c }
     \ell_1 & \ell_2 & \ell_3 \\
     0 & 0 & 0
  \end{array} \right) \label{sl3js}.
\een
\ah{The more usual skewness parameters are related to the skew-spectra by}:
 \be 
  S^{(i)} = {1 \over 4 \pi}\sum_{\ell}\Xi_{\ell}\; S^{(i)}_\ell.
\label{eq:S_l}
\ee
The bispectrum $B_{\ell_1\ell_2\ell_3}$ used here defines the three-point correlation function in the harmonic domain.
In general a reduced bispectrum $b_{\ell_1\ell_2\ell_3}$ is commonly used:
\be
\la \Psi_{\ell_1m_1}\Psi_{\ell_2m_2}\Psi_{\ell_3m_3} \ra_c = \left ( \begin{array}{ c c c }
     \ell_1 & \ell_2 & \ell_3 \\
     m_1 & m_2 & m_3
  \end{array} \right) B_{\ell_1\ell_2\ell_3}; 
\quad\quad B_{\ell_1\ell_2\ell_3} = I_{\ell_1\ell_2\ell_3}b_{\ell_1\ell_2\ell
_3}.
\label{eq:bispec_def}
\ee
This set of equations constitutes one of the main results in this paper.
$b_l$ represents the experimental beam $b_{\ell}\equiv b_\ell(\theta_s) = \exp\left \{-\ell(\ell+1)\sigma_b^2/2 \right \}$
with $\sigma_b = {\theta_s} /\sqrt{8 \ln 2}$ for a Gaussian beam. Each of these spectra probes the same bispectrum 
$B_{\ell\ell_1\ell_2}$ with 
different weights for individual triplets of modes 
that specifies the bispectrum $(\ell,\ell_1,\ell_2)$. Each triplets of modes specifies a triangle in the harmonic domain.
The skew-spectra sum over all possible configurations of the bispectrum keeping one of its sides $\ell$ fixed.
For each individual choice of $\ell$ we can compute the skew-spectra  $S_\ell^{(i)}$.

The expression for the optimum skew-spectrum estimator $S_\ell^{\rm opt}$ and its one-point
counterpart or the optimum skewness  $S^{\rm opt}$ are given by \citep{MuHe10}:
\ben
S_l^{\rm opt} = {1 \over 2\ell +1 }\sum_{\ell_1\ell_2} {\hat B_{\ell\ell_1\ell_2}B_{\ell\ell_1\ell_2} \over 
\myC_{\ell}\myC_{\ell_1}\myC_{\ell_2}}; \quad\quad S^{\rm opt} = \sum_{\ell}(2\ell+1) S_\ell^{\rm opt}.
\label{eq:muhe10}
\een
Here $\hat B_{\ell\ell_1\ell_2}$ is the bispectrum estimated from the data and $B_{\ell\ell_1\ell_2}$
is the theoretical model under consideration. The optimum estimator for various models that we consider
are presented in Figure \ref{fig:opt}.

An alternative is to formulate the analysis in real space, using the two-to-one {\em cumulant correlators} $S^{(i)}(\theta_{12})$:
\ben
 && S^{(0)}(\theta_{12}) \equiv {1 \over \sigma_0^4} \; S^{\Psi^2,\Psi} \;(\theta_{12}) 
\equiv  {1 \over \sigma_0^4}\la\Psi^2(\oh_1)\Psi(\oh_2)\ra 
= {1 \over 4\pi}\sum_l \Xi_{\ell} \; S^{(0)}_\ell P_\ell(\cos\theta_{12}); \label{eq:cumu1}\\ 
&& S^{(1)}(\theta_{12}) \equiv -{3 \over 4 \sigma_0^2\sigma_1^2} \;
S^{\Psi^2,\nabla^2\Psi}(\theta_{12}) \equiv  -{3 \over 4\sigma_0^2\sigma_1^2}\la\Psi^2(\oh_1)\nabla^2 \Psi(\oh_2)\ra = 
{1 \over 4 \pi}\sum_l \Xi_{\ell} \; S^{(1)}_\ell P_\ell(\cos\theta_{12}); \label{eq:cumu2}\\ 
 && S^{(2)}(\theta_{12}) \equiv -{3 \over \sigma_1^4} S^{\cb{\nabla\Psi\cdot\nabla\Psi,\nabla^2\Psi}}(\theta_{12})  \equiv  -{3 \over \sigma_1^4}\la\nabla\Psi(\oh_1)\cdot\nabla\Psi(\oh_1)\cb{\nabla^2}\Psi(\oh_2)\ra = 
 {1 \over 4 \pi}\sum_l \Xi_{\ell} \; S^{(2)}_\ell P_\ell(\cos\theta_{12}); \label{eq:cumu3}\\
&& S^{\rm opt}(\theta_{12}) \equiv  \sum_{\ell}(2\ell+1) S_\ell^{\rm opt} P_\ell(\cos\theta_{12}).
\een
Here, $\theta_{12}$ represents the angle of separation between $\oh_1$ and $\oh_2$, and $P_\ell$ is a Legendre polynomial. In the zero angular separation limit $\theta_{12}=0$ the two-point objects reduce to their one-point counterparts 
$ S^{(0)}(0)\equiv S^{(0)}$;
$S^{(1)}(0)\equiv S^{(1)}$ and $S^{(2)}(0)\equiv S^{(2)}$ respectively.
Similar construction is possible for the case of optimum estimator too.
In addition to the real-space description and its harmonic counterpart the needlet basis provides an intermediate choice.
We will consider the skew-spectra in needlet basis in \textsection\ref{sec:needlet}.

The extraction of skew-spectra
from data is relatively straightforward. It consists of construction the relevant maps in real space, either
by algebraic or differential operation, and then cross-correlating them in the multipole space.
The issues related to mask and noise will be dealt with in later sections. We will show that even in the 
presence of a mask the computed skew spectra can be inverted to give an unbiased estimate of all-sky skew-spectra,
with noise only affecting the scatter.

To derive the above expressions, we first express the spherical harmonic expansion of the fields 
$\nabla^2 \myf(\oh)$, $\nabla \myf(\oh) \cdot \nabla \myf(\oh)$ and $\myf^2(\oh)$ in terms
of the harmonics of the original fields, $\myf_{lm}$. These expressions involve the 3j functions
as well as factors that depend on various $\ell_i$-dependent weight factors. These aspects and
related issues have already been dealt with in previous publications in different contexts \citep{MuSmWaCo,MuSmJouCo}. 

We can the define the power spectrum associated with the MFs through the following third-order expression:
\be
{v}_k^{(2)}\cb{(\nu)} = \sum_\ell \Xi_{\ell}\;\cb{[{v}_k^{(2)}]}_\ell(\nu)  = {1 \over 6 } \sum_\ell \Xi_{\ell} \left \{ S^{(0)}_\ell {\cal H}_{k+2}(\nu) + {k \over 3 } S^{(1)}_\ell {\cal H}_{k}(\nu) +
{k(k-1) \over 6 } S^{(2)}_\ell {\cal H}_{k-2}(\nu) + \cdots \right \}.
\label{eq:nu_spec}
\ee
The three skewness parameters define the triplets of Minkowski Functionals. At the level of two-point statistics, in the harmonic
domain we have three power-spectra associated with Minkowski-Functionals
${v}_k^{(2)}$ that depend on the three skew-spectra 
we have defined. We will show later in this paper that the fourth-order correction terms too have a similar
form with an additional monopole contribution that can be computed from the lower-order one-point terms
such as the three skewness defined here. The result presented here is important and implies that we 
can study the contributions to each of the MFs $v_k(\nu)$ as a function of harmonic
mode $\ell$. This is a especially significant result as various forms of non-Gaussianity will have different $\ell$
dependence and hence they can potentially be distinguished. The ordinary MFs add contributions from
individual $\ell$ modes and hence have less power in differentiating various contributing sources of non-Gaussianity.
This is one of main motivations to extend the concept of MFs (single numbers) to one-dimensional objects similar
to the power spectrum. 

In Figure \ref{fig:S_loc} we have presented the three different skew-spectra for the local model as a function of 
the harmonics $\ell$. The skew-spectra for a generic bispectrum \cb{are} defined in Eq. (\ref{sl1}), 
Eq. (\ref{sl2}) and in
 Eq. (\ref{sl3}). The skew spectra are sensitive to the smoothing $b_\ell$ moreover the skew-spectra 
at a given $\ell$ depend on the bispectrum defined over the entire range of $\ell$ being probed.
The skew-spectra for equilateral, orthogonal and enfolded model are presented in Figure \ref{fig:S_eq},
Figure \ref{fig:S_or} and Figure \ref{fig:S_en} respectively.
The normalization parameters for these plots are set to be equal to unity; i.e. we take $f_{\rm NL}^{\rm loc}=1$ and similarly for other models. The skew-spectra will scale linearly with these $f_{\rm NL}$ parameters. \cb{In addition to the amplitude of the skew-spectra, comparing the figures we can see that
equilateral model produced the most distinct type of skew-spectra which is very different from
all other models. The skew-spectra of other models too have very different signature
especially at high $\ell$s.}

\cb{For our computation of the MFs, we have used the freely-available software archive
SHTOOLS\footnote{shtools.ipgp.fr}. In particular, we used its Wigner-3j symbol
routine that provides accurate numerical convergence, especially for high values of $\ell$. We used a parallel implementation 
of Eq. (\ref{sl1})-Eq. (\ref{sl3})
for evaluation of the skew-spectra. For the separable models of bispectra considered 
here two hours of computations were required on 20 CPUs. The computations were 
dominated by evaluation of $3j$ symbols.}

\cb{The unresolved point sources are mostly unresolved galaxies, i.e. radio galaxies not emitting strongly 
enough for their individual detection; which emit in radio frequencies via the synchrotron process
or dusty starburst galaxies which are observed via thermal emission of their dust.} The integrated 
diffuse emission from all galaxies constitute what is commonly known as the Cosmic Infrared Background (CIB).
The brightest point sources can be removed using an appropriate mask. 
The unresolved point sources, however, do contribute to the CMB bispectrum, \ah{and in general they will be clustered and require more detailed modelling}.  The skew-spectra
for Poisson-distributed point sources are plotted in Figure \ref{fig:S_ps}. The specific model for the bispectrum that we 
consider is given in Eq. (\ref{eq:point}), and the point sources are expected to dominate at higher $\ell$
values. The normalization for the point source bispectrum is set by the parameter $b_{\rm PS}$.  \cb{Assuming
point sources are not clustered and can be represented as a Poissionian distribution we can write
the corresponding bispectrum as:}
\be
b_{\ell_1\ell_2\ell_3} = b_{\rm PS}.
\label{eq:point}
\ee
The exact value of  the amplitude $b_{\rm PS}$ depends on the limiting flux used in a specific survey. In our
study we have taken $b_{\rm PS}=\cb{10^{-29}}\mu {\rm K}^3$. The results of our computations are plotted in Figure \ref{fig:S_ps}.
We have considered three different Gaussian beams as indicated $\theta_s = 10'$, $25'$, 
$50'$, and $100'$. 
The skew-spectra $S^{(2)}_\ell$, which puts more weights on smaller angular scales, is
more dominated by point-source contributions. The cumulant correlators introduced in Eq. (\ref{eq:cumu1})-Eq. (\ref{eq:cumu3}) 
are depicted in Figure \ref{fig:cumu_loc} for four different models that we have considered.

Next, we consider the higher-order corrections to the MFs. These corrections take contributions from the
trispectrum. Corrections to the individual MFs can be expressed in terms of a set of four kurtosis terms which are formed 
from the trispectra. These kurtosis terms are one-point estimators and they differ in the way they sample the 
individual modes of the trispectra defined by the quadruplet of harmonic number $\{\ell_i\}; i=1,2,3,4$. 
These generalized kurtosis parameters (denoted by $K^{(i)}$) can be generalized to {\em kurt-spectra}, denoted as $K^{(i)}_\ell$, 
in a manner very similar to the skew-spectra.
These kurt-spectra can be used to express the next-order corrections to the power spectra 
associated with MFs. 
\section{Estimators and their Covariance}
\label{sec:estim}
The results derived above correspond to the all-sky and no-noise situation.
However, in reality often we have to deal with issues that are related to the presence of a mask and (inhomogeneous) noise.
To correct for the effect of a mask and the noise we will follows the pseudo-$\myC_{\ell}$ method
devised by \cite{Hiv} for power spectrum analysis and later developed by \cite{MuX09} for analyzing the skew spectra
and kurt-spectrum \citep{Mu_kurt10}. 

The partial sky coverage introduces mode-mode coupling in the harmonic domain and individual masked harmonics
become linear combinations of all-sky harmonics. The coefficients for this linear transformation depend on 
the mask through its harmonic coefficients. We will devise a method that
can be used to correct for the mode-mode coupling. If we have a generic field
$A(\oh)$ and $B(\oh)$ we denote their harmonic decomposition in the presence of a mask $w(\oh)$
as $\tilde A_{\ell m}$ and $\tilde B_{\ell m}$. Notice that the mask is completely general and 
our results do not depend on any specific symmetry requirements such as the azimuthal symmetry.
The fields $A$ and $B$ may correspond to any of the fields we have considered above. In a generic
situation $A$ and $B$ will denote composite fields and  the harmonics $A_{\ell m}$ and $B_{\ell m}$
will correspond to any of the harmonics used in Eq. (\ref{sl1})-Eq. (\ref{sl3}).
\ben
&& \tilde A_{\ell m} = \int ~d\oh Y^*_{\ell m}(\oh)~ w(\oh)~ A(\oh); \quad\quad  \tilde B_{\ell m} = \int ~d\oh~ Y^{*}_{\ell m}(\oh)~w(\oh) B(\oh); \\
&& \tilde A_{\ell m} = \sum_{\ell_i m_i} (-1)^m~I_{\ell\ell_1\ell_2} \left ( \begin{array}{ c c c }
     \ell_1 & \ell_2 & \ell \\
     m_1 & m_2 & -m
  \end{array} \right) w_{\ell_1m_1} A_{\ell_2m_2}; \quad
 \tilde B_{\ell m} = \sum_{\ell_im_i} (-1)^m I_{\ell\ell_1\ell_2}\left ( \begin{array}{ c c c }
     \ell_1 & \ell_2 & \ell \\
     m_1 & m_2 & -m
  \end{array} \right) w_{\ell_1m_1} B_{\ell_2m_2}.
\een
The above expression relates the masked harmonics denoted by $\tilde A_{\ell m}$ and $\tilde B_{\ell m}$  with their all-sky counterparts
$A_{\ell m}$ and $B_{\ell m}$ respectively. In their derivation we use the Gaunt integral to express the overlap integrals
involving three spherical harmonics in terms of the $3j$ symbols \citep{Ed68}.
The expressions also depend on the harmonics of the mask $w_{\ell m}$. If we now denote the
(cross) power spectrum constructed from the masked harmonics and denote it by $\tilde S_\ell$ and its all-sky counterpart
by $S_\ell$ we can write:
\be
\tilde S_\ell^{A,B} = {1 \over \Xi_{\ell}} \sum_{m} \tilde A_{\ell m} \tilde B^*_{\ell m}; \quad
\tilde S_\ell^{A,B} = \sum_{\cb{\ell'}} {\mathbb M}_{\ell\ell'} S_\ell^{A,B}; \quad {\mathbb M}_{\ell\ell'} = {1 \over \Xi_{\ell}}\sum_{\cb{\ell''}} I^2_{\ell\ell'\ell''} |w_{\ell''}|^2; \quad \left \{ A,B \right \} \in \left \{\myf,\myf^2, (\nabla \myf\cdot \nabla \myf), \nabla^2 \myf \right\}.  \\
\label{eq:alm_est}
\ee
\n
\cb{Here $w_\ell$ represents the power spectrum of the mask $w(\oh)$ i.e. $w_\ell = {1 \over 2\ell+1}\sum_{m=-\ell}^{m=\ell}w_{\ell m}w^*_{\ell m}$}.

In the above derivation we have used the orthogonality properties of the 3j symbols. It is interesting to
notice that the {\em convolved} power spectrum estimated from the masked sky is 
a linear combination of all-sky spectra and depends only on the power spectra of the mask used.
The linear transform is encoded in the mode-mode coupling matrix ${\mathbb M}_{\ell\ell'}$ which is constructed
from the knowledge of the power spectrum of the mask. In certain situations
where the sky coverage is low the direct inversion of the mode mixing matrix ${\mathbb M}$ may not be possible \cb{due} to its singularity and
binning may be essential. \cb{Based} on these results it is possible to define an unbiased estimator that we denote by $\hat S_\ell^{A,B}$.
The noise due to its Gaussian nature, do not contribute in these estimators which remain unbiased. However, the
presence of noise is felt in an increase in the scatter or covariance of these estimator\cb{s} which can be computed analytically:
\be
\hat S_\ell^{A,B} = \sum_{\cb{\ell'}} [{\mathbb M}^{-1}]_{\ell\ell'} \tilde S_{\ell'}^{A,B}; \quad  
\la \delta \hat S_\ell^{A,B} \delta \hat S_{\ell'}^{A,B} \ra =  \sum_{LL'}{\mathbb M}^{-1}_{\ell L} \la \delta \tilde S_{L}^{A,B} \delta \tilde S_{L'}^{A.B} \ra {\mathbb M}^{-1}_{L'\ell'};
\quad \langle \hat S_\ell^{A,B} \rangle = S_\ell^{A,B}; \quad \left \{ A,B \right \} \in \left \{\myf,\myf^2, (\nabla \myf\cdot \nabla \myf), \nabla^2 \myf \right\}.  \\
\label{eq:auto_cov}
\ee
Notice that the mode-coupling matrix ${\mathbb M}$ is independent of the particular choice of the skew-spectrum.
Hence the same coupling matrix can be used to extract the power spectrum associated with the MFs,
as the MFs are constructed from the linear combinations of generalized skew-spectra.
\be
[\hat {\cal V}_k^{(2)}]_\ell =  \sum_{\ell'} [{\mathbb M}^{-1}]_{\ell\ell'} [\tilde {\cal V}_k^{(2)}]_\ell; \quad \quad
\la \delta \hat {\cal V}_k^{(2)} \delta \hat {\cal V}_{k'}^{(2)} \ra =  
\sum_{LL'}{\mathbb M}^{-1}_{\ell L} \la \delta [\tilde {\cal V}_{k}^{(2)}]_\ell \delta [\tilde {\cal V}_{k'}^{(2)}]_{\ell'} \ra {\mathbb M}^{-1}_{L'\ell'}. 
\ee
The variance $\la \delta S_\ell^{A,B} \delta S_{\ell}^{A,B} \ra$ of various estimators can be constructed using the following procedure:  
\ben
&& \la \delta S_\ell^{A,B} \delta S_{\ell}^{A,B} \ra =   {f^{-1}_{\rm sky} \over \Xi_{\ell}} \left [\myC_\ell^{A,A}\myC_\ell^{B,B} + [S_\ell^{A,B}]^2 \right ]; \\
&& \myC_\ell^{\nabla\Psi\cdot\nabla\Psi,\nabla\Psi\cdot\nabla\Psi} = \sum_{\cb{\ell_1\ell_2}} \; \Xi_{\ell}\; \myC^{}_{\ell_1} \myC^{}_{\ell_2}  
[{\Pi}_{\ell_1}+{\Pi}_{\ell_2}-{\Pi}_{\ell}]^2  
J^2_{\ell_1\ell_2\ell}; \quad \\
&& \myC_{\ell}^{[\myf^2, \myf^2]} = \sum_{\ell_1\ell_2} \; \Xi_{\ell}\; \myC^{}_{\ell_1} \myC^{}_{\ell_2} J^2_{\ell_1\ell_2\ell};\quad
 \myC_{\ell}^{[\nabla^2 \myf, \nabla^2 \myf]} = {\Pi}_{\ell}^2\;\myC_\ell; \quad \myC_{\ell}^{\Phi,\Phi} \equiv \myC_\ell.
\label{eq:factor1}
\een
\n
\ah{We have used standard relations of 3j symbols, summarised in Appendix \ref{3j}, to derive these results.}
$\myC_\ell^{A,A}$ denotes the power spectrum of a generic map $A(\oh)$ 
that is used for the construction of generalized skew-spectra and $f_{\rm sky}$ is the fraction of sky coverage.
 The derivation depends on a Gaussian approximation i.e. we ignore higher-order non-Gaussianity in the fields. 
$\myC_\ell$ is the ordinary CMB power spectra, including the effect of instrumental noise, $\myC_\ell = \myC^{\rm S}_\ell + \myC_\ell^{\rm N}$. The first term represents cosmic variance and the second term is the effect of instrumental noise.
For a survey with homogeneous noise, ignoring the effect of the beam we can write $\myC_\ell^N = \Omega_{\rm p} \sigma_{\rm N}^2$ where 
$\Omega_p$ is the pixel area and $\sigma^2_{\rm N}$ is the noise variance. In a noise-dominated regime the
MFs can be approximated by a Gaussian. The explicit expressions for the three skew-spectra that we have considered are as follows: 
\ben
&& \quad\quad \la [\delta S_\ell^{\Psi^2,\Psi}]^2 \ra = { f^{-1}_{\rm sky} \over \Xi_{\ell} }
\left [ [S_\ell^{\Psi^2,\Psi}]^2 + \myC^{\Psi^2,\Psi^2}_{\ell}\myC^{\Psi,\Psi}_{\ell} \right]; \label{eq:scat1}\\
&& \quad\quad  \la [\delta S_\ell^{\Psi^2,\nabla^2\Psi}]^2 \ra = { f^{-1}_{\rm sky} \over \Xi_{\ell} }
\left [ [S_\ell^{\Psi^2,\nabla^2\Psi}]^2 + \myC^{\Psi^2,\Psi^2}_{\ell}\myC^{\nabla^2\Psi,\nabla^2\Psi}_{\ell} \right]; \label{eq:scat2}\\
&& \quad\quad \la [\delta S_\ell^{\nabla\Psi\cdot\nabla\Psi,\nabla^2\Psi} ]^2 \ra = { f^{-1}_{\rm sky} \over \Xi_{\ell} }
\left [ [S_\ell^{\nabla\Psi\cdot\nabla\Psi,\nabla^2\Psi}]^2 + \myC^{\nabla\Psi\cdot\nabla\Psi,\nabla\Psi\cdot\nabla\Psi}_{\ell}\myC^{\nabla^2\Psi,\nabla^2\Psi}_{\ell} \right]. \label{eq:scat3}
\een
The estimators for various skew-spectra are expected to be correlated to a certain extent. These can be expressed using 
following expression: 
\be
\la \delta S_\ell^{A_1,B_1} \delta S_{\ell}^{A_2,B_2} \ra =  {f^{-1}_{\rm sky} \over \Xi_{\ell}} \left [\myC_\ell^{A_1,A_2}\myC_\ell^{B_1,B_2} +S_\ell^{A_1,B_2}S_\ell^{A_2,B_1}\right ]; \quad
 \left \{ A_1,B_1,A_2,B_2 \right \} \in \left \{\myf,\myf^2, (\nabla \myf\cdot \nabla \myf), \nabla^2 \myf \right\}.\\
\label{eq:cross_cov}
\ee
The above results are sufficient to compute the lowest-order corrections to MFs 
due to the presence of non-Gaussianity, as well as the scatter in the estimates in the presence of realistic mask and noise.
The explicit expressions are: 
\ben
&& \quad\quad \la \delta S_\ell^{\Psi^2,\Psi} \delta S_{\ell}^{\Psi^2,\nabla^2\Psi} \ra = { f^{-1}_{\rm sky} \over \Xi_{\ell} }
\left [ S^{\Psi^2,\Psi}_{\ell}S^{\Psi^2,\nabla^2\Psi}_{\ell} +
 \myC_\ell^{\Psi^2,\Psi^2}\myC_{\ell}^{\Psi,\nabla^2\Psi}\right]; \label{eq:cross1}\\
&& \quad\quad  \la \delta S_\ell^{\Psi^2,\Psi} \delta S_{\ell}^{\nabla\Psi\cdot\nabla\Psi,\nabla^2\Psi} \ra = { f^{-1}_{\rm sky} \over \Xi_{\ell} }
\left [ S_\ell^{\Psi^2,\nabla^2\Psi} S_\ell^{\Psi,\nabla\Psi\cdot\nabla\Psi} + \myC^{\Psi^2,\nabla\Psi\cdot\nabla\Psi}_{\ell}\myC^{\Psi,\nabla^2\Psi}_{\ell} \right]; \label{eq:cross2}\\
&& \quad\quad \la \delta S_\ell^{\nabla\Psi\cdot\nabla\Psi,\nabla^2\Psi} \delta S_{\ell}^{\Psi^2,\nabla^2\Psi}  \ra = { f^{-1}_{\rm sky} \over \Xi_{\ell} }
\left [ S_\ell^{\nabla\Psi\cdot\nabla\Psi,\nabla^2\Psi} S_\ell^{\Psi^2,\nabla^2\Psi} + 
\myC^{\nabla\Psi\cdot\nabla\Psi,\Psi^2}_{\ell}\myC^{\nabla^2\Psi,\nabla^2\Psi}_{\ell} \right] \label{eq:cross3}.
\een
The additional cross-spectra that are introduced above are:
\be
\myC_{\cb{\ell}}^{\Psi,\nabla^2\Psi}=-\Pi_{\ell}\myC_\ell; \quad\quad \myC_\ell^{\Psi^2,\nabla\Psi\cdot\nabla\Psi}=2\sum_{\cb{\ell_1\ell_2}} 
\;\Xi_{\ell}\;\myC^{}_{\ell_1} \myC^{}_{\ell_2} 
[{\Pi}_{\ell_1}+{\Pi}_{\ell_2}-{\Pi}_{\ell}] J^2_{\ell_1\ell_2\ell}. \quad
\ee
The signal-to-noise ratio and the cross-correlation coefficients among various skew-spectra are defined as:
\be
\left ({S \over N}\right )^{(i)}_\ell  = {S^{(i)}_{\ell} \over \la [\delta S^{(i)}_{\ell}]^2 \ra^{1/2} }; \quad\quad r^{ij}_{\ell} = {\la \delta S^{(i)}_{\ell} \delta S^{(j)}_{\ell}\ra \over \la [\delta S^{(i)}_{\ell}]^2\ra^{1/2}
\la [\delta S^{(j)}_{\ell}]^2 \ra^{1/2}}.
\label{eq:corrij}
\ee

To compute the cross-correlation of skew-spectra $S_{{\rm I},\ell}^{A_1,B_1}$ and $S_{{\rm II},\ell}^{A_1,B_1}$, which
source different bispectra $B_{{\rm I},\ell_1\ell_2\ell_3}$ and $B_{{\rm II},\ell_1\ell_2\ell_3}$ (either primary and secondary or two different models of primary
bispectra) from the same data, we can use the following simple extension of Eq. (\ref{eq:cross_cov}):
\be
\la \delta S_{{\rm I},\ell}^{A_1,B_1} \delta S_{{\rm II},\ell}^{A_2,B_2} \ra =  
{f^{-1}_{\rm sky} \over \Xi_{\ell}} \left [\myC_\ell^{A_1,A_2}\myC_\ell^{B_1,B_2}
 +\sqrt{S_{{\rm I},\ell}^{A_1,B_2}S_{{\rm II},\ell}^{A_2,B_1}}\sqrt{S_{{\rm II},\ell}^{A_1,B_2}S_{{\rm I},\ell}^{A_2,B_1}}\right ]; \quad
 \left \{ A_1,B_1,A_2,B_2 \right \} \in \left \{\myf,\myf^2, (\nabla \myf\cdot \nabla \myf), \nabla^2 \myf \right\}.\\
\label{eq:cross_cov_gen}
\ee
These results will be valid for near all-sky coverage and in a regime where noise dominates. Bias from inaccurate
foreground subtraction is ignored. The results presented here can be extended to include estimation of kurt-spectra. 

\begin{figure}
\begin{center}
{\epsfxsize=16.5 cm \epsfysize=5.25 cm {\epsfbox[25 521 583 715]
{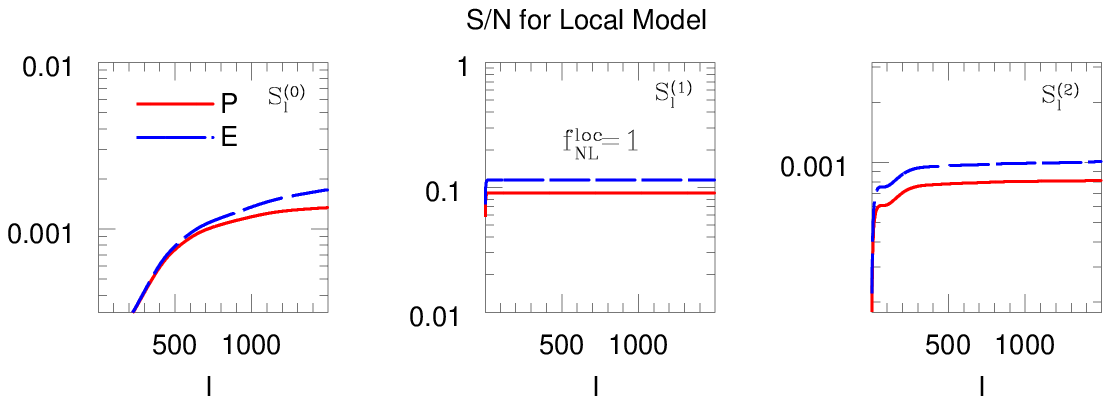}}}
\end{center}
\caption{The cumulative signal-to-noise ratio,
$\sum_\ell ({\rm S}/{\rm N})^{\rm loc}_\ell$, for various skew-spectra that correspond to the {\em local}
model of primordial non-Gaussianity.
These results correspond to $f^{\rm loc}_{\rm NL}=1$.
Each panel shows results for  Planck (143 GHz channel) and for Epic (150 GHz channel). 
The left, middle and right panels correspond to $S^{(0)}_{\ell}$, $S^{(1)}_{\ell}$ and $S^{(2)}_{\ell}$ respectively.
The expressions for the covariances are listed in Eq. (\ref{eq:scat1})-Eq. (\ref{eq:scat3}). 
We have assumed a full sky coverage $f_{\rm sky}=1$ for both of these experiments.}
\label{fig:scatter_loc}
\end{figure}
\begin{figure}
\begin{center}
{\epsfxsize=16.5 cm \epsfysize=5.25 cm {\epsfbox[25 521 583 715]
{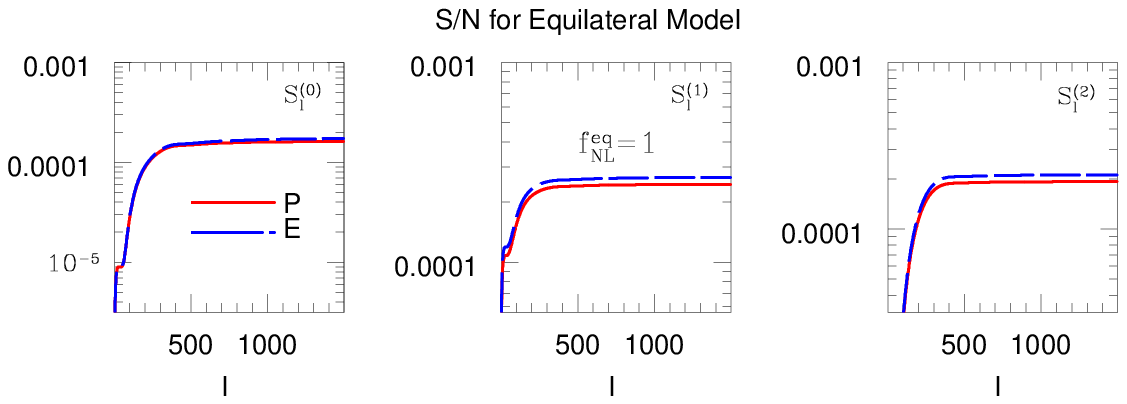}}}
\end{center}
\caption{Same as previous figure but for the {\em equilateral} model ($f^{eq}_{\rm NL}=1$) of primordial non-Gaussianity. }
\label{fig:scatter_eq}
\end{figure}
\begin{figure}
\begin{center}
{\epsfxsize=16.5 cm \epsfysize=5.25 cm {\epsfbox[25 521 583 715]
{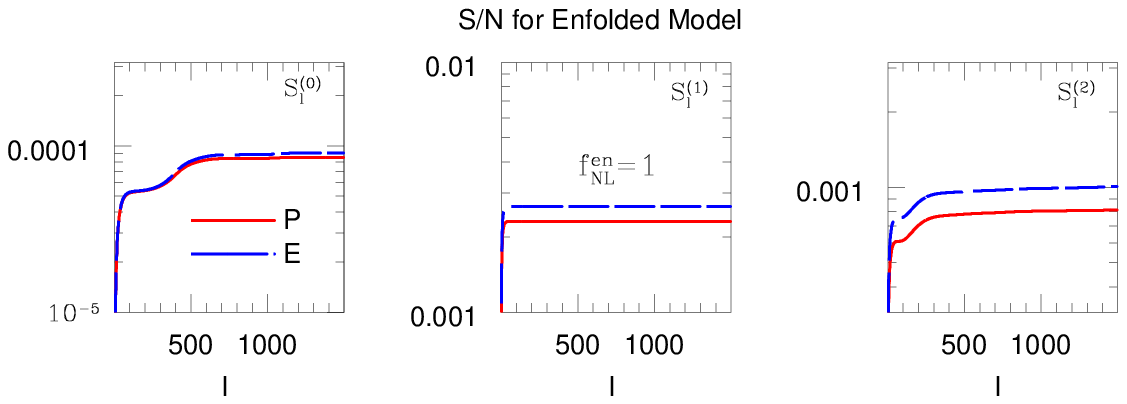}}}
\end{center}
\caption{Same as previous figure but for the {\em enfolded} model ($f^{en}_{\rm NL}=1$)  of primordial non-Gaussianity.}
\label{fig:scatter_en}
\end{figure}
\begin{figure}
\begin{center}
{\epsfxsize=16.5 cm \epsfysize=5.25 cm {\epsfbox[25 521 583 715]
{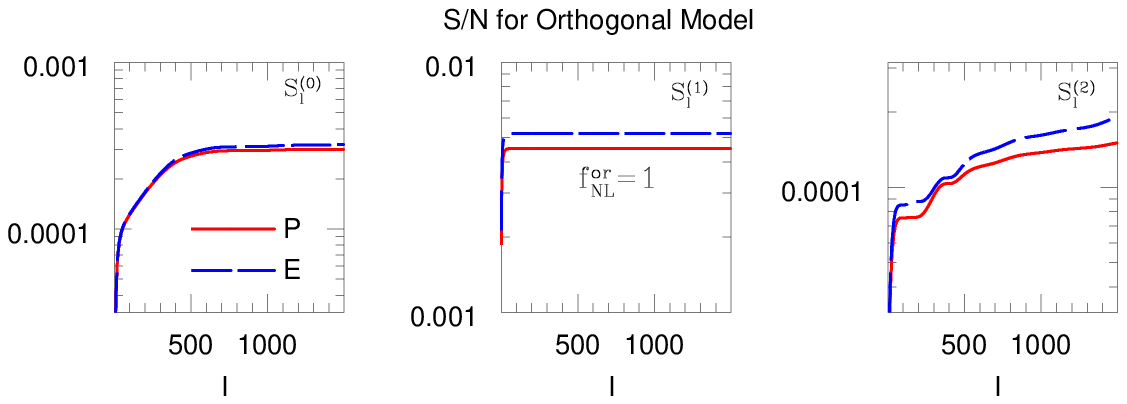}}}
\end{center}
\caption{Same as previous figure but for the {\em orthogonal} model ($f^{orth}_{\rm NL}=1$) of primordial non-Gaussianity.}
\label{fig:scatter_orth}
\end{figure}
\cb{The numerical results for the ${\rm S/N}$ for the local, equilateral, enfolded and orthogonal models 
are plotted in Figs. \ref{fig:scatter_loc} - \ref{fig:scatter_orth} respectively.
From these results, 
we find that in most cases the $\rm{S/N}$ is dominated by
$\ell < 500$. Among the three skew-spectra we have considered $S_\ell^{(1)}$ achieves 
the maximum $\rm{S/N}$ due to optimum $\ell$ weighting. This is in agreement with our previous 
studies of MFs in weak lensing \citep{MuSmWaCo}, thermal SZ \citep{MuSmJouCo} studies.
Individual $S_\ell$s differ in their $\ell$-dependent weightings of the bispectrum, with the
weights for $S^{(1)}_{\ell}$ appearing to give the optimum balance among the three $S_{\ell}$ considered.
The increase in $\rm{S/N}$ by changing experimental set up from Planck to Epic is nominal
as most of the signal is at low $\ell$.} \ah{We see that in order to reach S/N$>3$ would require $f_{\rm NL}>20,3 \times 10^3,60,10^3$ for the local, equilateral, orthogonal and enfolded models respectively.} 
\section{Modal decomposition and reconstructing Minkowski Functionals}
\label{sec:modal}
In recent years modal decomposition of a generic bispectrum, in terms of a separable or orthogonal basis, has been proposed (e.g. \cite{FLS10}).
Three-dimensional modes ${\cal Q}_n^{\ell_1\ell_2\ell_3}(x)$  (to be defined later) are constructed from one-dimensional modes $q^{\ell}_p(x)$, and
the coefficients in the expansion can then be used to reconstruct the bispectrum or trispectrum.
The primary aim of this section is to express the skewness-spectra introduced in the text in terms of the coefficients characterizing the
modal expansion of the bi- and trispectrum.

Following the procedure detailed in \cite{FLS10} we introduce the following modal decomposition of the reduced bispectrum
$b_{\ell_1\ell_2\ell_3}$ in terms of modal function denoted as ${\cal Q}_n^{\ell_1\ell_2\ell_3}(x)$:
\ben
&& q^{\ell}_p(x) = {2 \over \pi }\int dk \; q_p(k) \; \Delta_{\ell}(k)\; j_{\ell}(kx); \quad\quad
{\cal Q}_n^{\ell_1\ell_2\ell_3}(x) \equiv  q_{\{p}^{\ell_1}(x) q_q^{\ell_2}(x) q_{r\}}^{\ell_3}(x);\\
&& b_{\ell_1\ell_2\ell_3}\equiv\Delta^2_{\Phi} f_{\rm NL} \sum_{n} \alpha_n^{\cal Q} \int x^2 dx  \; {\cal Q}_n^{\ell_1\ell_2\ell_3}(x).
\label{eq:bispec_mode}
\een
The separable basis functions $q_p^{\ell}$ are convolutions of the spatial basis functions $q_p(k)$ and the transfer function $\Delta_{\ell}(k)$.
They reduce the dimensionality of the integral by expressing the three-dimensional
integral in terms of three one-dimensional integrals that are easy to evaluate. The three-dimensional basis
functions ${\cal Q}_n^{\ell_1\ell_2\ell_3}(x)$ 
are then constructed from the one-dimensional $q_{p}^{\ell}(x)$ basis functions. The curly brackets in  $q_{\{p}^{\ell_1}(x) q_q^{\ell_2}(x) q_{r\}}^{\ell_3}(x)$ 
represents all possible permutations of the indices $\ell_1\ell_2\ell_3$. The index $n$  
represents a specific combination of one-dimensional modes described by the triplets of indices$\{ {pqr}\}$; a mapping between the two
is implicitly assumed $n\leftrightarrow\{pqr\}$ below.

The speed and accuracy of the modal decomposition depends ultimately on the smoothness of the reduced bispectrum.
The modal decomposition above was carried out on separable basis functions $q_p^\ell(x)$. Nevertheless, the Gram-Schmidt orthogonalisation
procedure can be employed to construct a set of orthogonal modes in which an equivalent analysis can be formulated.
Completeness of the orthonormal basis is important for the accuracy of the modal decomposition.  \ah{The bispectrum is expanded in a finite set of modes:}
\ben
&& {v_{\ell_1}v_{\ell_2}v_{\ell_3}\over \sqrt{\myC_{\ell_1}\myC_{\ell_2}\myC_{\ell_3}}} b_{\ell_1\ell_2\ell_3} = 
\sum_{n \leftrightarrow \{pqr\}} \alpha^{\cal Q}_n {\cal Q}_n; \\
&& M_p(\oh) = \sum_{lm} q_p^\ell(x) {\Psi_{\ell m} \over v_{\ell} \sqrt{\myC_{\ell}}} Y_{\ell m}(\oh); \quad\quad
M_{n}(\oh) = M_{p}(\oh) M_{r}(\oh) M_{s}(\oh).
\een
Here $M_p(\oh)$ are filtered maps constructed from the harmonics $\Psi_{\ell m}$ of the original map $\Psi$. The functions
$v_{\ell}= (2\ell+1)^{1/6}$ are introduced to remove any residual $\ell$-dependence in the reduced bispectrum $b_{\ell_1\ell_2\ell_3}$. \ah{We also define:}
\ben
&& \la \beta_n^{\cal Q}\ra \equiv \int d\oh \int x^2 dx \; \la M_{n}(\oh)\ra  = \sum_{\cb{n'}} \Gamma_{nn'} \alpha_{n'}^{\cal Q}; 
\quad\quad  \hat\alpha_n^{\cal Q} = \sum_n [\Gamma^{-1}]_{nn'} \beta_{n'}^{\cal Q}; \quad\quad
\Gamma_{nn'} = \sum_{\ell_1\ell_2\ell_3} {w_{\ell_1\ell_2\ell_3} \over v^2_{\ell_1}v^2_{\ell_2}v^2_{\ell_3}} {\cal Q}_n {\cal Q}_{n'}.
\een
Next the triplets of skew-spectra $S^{(0)}_{\ell},S^{(1)}_{\ell},S^{(2)}_{\ell}$ can be constructed from the modal coefficients $\alpha_n^{\cal Q}$:
\ben
&& 3\sigma_0^4\; \hat S_{\ell}^{(0)} \equiv \sum_{\ell_1\ell_2} I^2_{\ell_1\ell_2\ell} b_{\ell_1\ell_2\ell} = \Delta^2_{\Phi}f_{\rm NL} \sum_n \sum_{\ell_1\ell_2} \hat \alpha_n^{\cal Q}
\int x^2 dx \; \;{\cal Q}_n^{\ell_1\ell_2\ell}(x); \quad\quad\\
&& 4\sigma_0^2\sigma_1^2 \; \hat S_{\ell}^{(1)} \equiv \sum_{\ell_1\ell_2} I^2_{\ell_1\ell_2\ell} (\Pi_{\ell_1}+\Pi_{\ell_2}+\Pi_{\ell_3})b_{\ell_1\ell_2\ell} = \Delta^2_{\Phi}f_{\rm NL} \sum_n\sum_{\ell_1\ell_2} \hat \alpha_n^{\cal Q}\;
(\Pi_{\ell_1}+\Pi_{\ell_2}+\Pi_{\ell})\int x^2 dx \; \;{\cal Q}_n^{\ell_1\ell_2\ell}(x); \quad\quad\\
&& 2 \sigma_1^4 \; \hat S_{\ell}^{(2)} \equiv \sum_{\ell_1\ell_2} I^2_{\ell_1\ell_2\ell} 
[(\Pi_{\ell_1}+\Pi_{\ell_2}-\Pi_{\ell})\Pi_{\ell}+{\rm cyc.perm.}]b_{\ell_1\ell_2\ell} \nn \\
&& \quad\quad\quad\quad = \Delta^2_{\Phi}f_{\rm NL} \sum_n\sum_{\ell_1\ell_2} 
\hat \alpha_n^{\cal Q} 
 [(\Pi_{\ell_1}+\Pi_{\ell_2}-\Pi_{\ell})\Pi_{\ell} +{\rm cyc.perm.}] \; 
\int x^2 dx {\cal Q}_n^{\ell_1\ell_2\ell}(x).
\label{eq:bispec}
\een
We have suppressed the experimental beams in these expressions for clarity.

Similar expressions for modal decomposition of the (reduced) \cb{trispectrum} can be found in \cite{RSF10}.
The modal expansion of the \cb{trispectrum} $\tau^{\ell_1\ell_2}_{\ell_3\ell_3}(\ell)$ requires a five-dimensional basis 
and the coefficients of expansion can be used to reconstruct the MFs.
\ben
&& \tau^{\ell_1\ell_2}_{\ell_3\ell_4}(\ell) \propto \Delta^3_{\phi} \sum_m \alpha_m^{\cal Q}\int x_1^2 dx_1 x_2^2 dx_2
{\cal Q}_{\ell_1\ell_2\ell_3\ell_4\ell}(x_1,x_2); \quad {\cal Q}_{\ell_1\ell_2\ell_3\ell_4\ell}(x_1,x_2)= q_p^{\ell_1}(x_1)q_r^{\ell_2}(x_1)q_s^{\ell_3}(x_2)q_{u}^{\ell_4}(x_2)r^{\ell}_v(x_1,x_2); \\
&& r_v^{\ell}(x_1,x_2) = {2 \over \pi} \int dk \,k\, r_v(k)\, j_{\ell}(kx_1)\,j_{\ell}(kx_2).
\label{eq:tripsec_mode}
\een
These expressions are useful for the construction of an estimator for $\tau$  from a CMB map which can then
be used in 
Eq. (\ref{eq:kurt_first})-Eq. (\ref{eq:kurt_spectra2}) for estimation of kurt-spectra associated with MFs.
This will provide a consistency check for the results obtained using direct estimators of MFs defined in Eq. (\ref{eq:direct1})
-Eq. (\ref{eq:kurt_spectra}).  

The actual data analysis pipelines for MF analysis and that of 3D modal decomposition are
very different. Relating perturbative expansion of MFs and using the modal decomposition 
to reconstruct the MFs at intermediate steps may lead to a better understanding of the
systematics affecting their estimation. 

\section{Odd-Parity Bispectrum and Minkowski Functionals}
\label{sec:odd}
Most analyses of the bispectrum assume the bispectrum to be of even parity. Recently the possibility 
of odd-parity bispectrum was underlined by \cite{KS11}. Such a bispectrum {\em cannot} arise from
projecting the 3D density perturbations. Nevertheless, the odd-parity bispectrum can result from
lensing of the CMB by a chiral gravitational wave background or from cosmological birefringence 
\citep{Komatsu2011,Feng06,Wu09}. Models
with a time-dependent quintessence field that couples to pseudo-scalar of the electromagnetic 
field and induced rotation of linear polarization and generate magnetic
``B'' mode polarization from pure electric ``E'' mode and hence
induce a parity odd mixed temperature-polarization bispectra \citep{Caroll98,CFJ90,LWM90}.

The reduced bispectrum $b_{\ell_1\ell_2\ell_3}$ introduced in the Eq. \cb{(\ref{eq:bispec_def})} is 
replaced by the following equation:
\ben
&& \la a_{\ell_1m_1}a_{\ell_2m_2}a_{\ell_3m_3}\ra = B_{\ell_1\ell_2\ell_3}\left ( \begin{array}{ c c c }
     \ell_1 & \ell_2 & \ell_3 \\
     m_1 & m_2 & m_3
  \end{array} \right); \quad B_{\ell_1\ell_2\ell_3}={\mathcal I}_{\ell_1\ell_2\ell_3}b_{\ell_1\ell_2\ell_3};\\
&& {\mathcal I}_{\ell_1\ell_2\ell_3} =  {{\Pi_{\ell_2}\Pi_{\ell_3}} \over \Pi_{\ell_1} -\Pi_{\ell_2} -\Pi_{\ell_3}}
\sqrt{\Xi_{\ell_1}\Xi_{\ell_2}\Xi_{\ell_3}\over 4 \pi}
\left ( \begin{array}{ c c c }
     \ell_1 & \ell_2 & \ell_3 \\
     0 & -1 & 1
  \end{array} \right).
\label{eq:odd_bi}
\een
It can be shown that ${\cal I}_{\ell_1\ell_2\ell_3} = I_{\ell_1\ell_2\ell_3}$ for even parity i.e.
$\ell_1+\ell_2+\ell_3={\rm even}$ but it remains non-zero also for odd-parity i.e. $\ell_1+\ell_2+\ell_3={\rm odd}$.
\ben
&& S_{\ell}^{(0)} = {2 \over 3 \sigma_0^4} \; {1\over \Xi_{\ell}}\sum_{\ell_1 >\ell_2}\;b_{\ell_1\ell_2\ell}\; {\mathcal I}^2_{\ell_1\ell_2\ell}\,b_{\ell_1}\,b_{\ell_2}\,b_{\ell}; \label{eq:odd1}\\
&& S_{\ell}^{(1)} = {1 \over 2 \sigma_0^2\sigma_1^2}\; {1\over \Xi_{\ell}}
\sum_{\ell_1 >\ell_2}(\Pi_{\ell_1}+\Pi_{\ell_2}+\Pi_{\ell})\;b_{\ell_1\ell_2\ell}\; {\mathcal I}^2_{\ell_1\ell_2\ell}\,b_{\ell_1}\,b_{\ell_2}\,b_{\ell}; 
\label{eq:odd2}\\
&& S_{\ell}^{(2)} = {1 \over \sigma_1^4} \; {1\over \Xi_{\ell}}
\sum_{\ell_1 >\ell_2}[(\Pi_{\ell}+\Pi_{\ell_1}-\Pi_{\ell_2})\Pi_{\ell_2} +{\rm cyc.perm.}]\;b_{\ell_1\ell_2\ell}\; {\mathcal I}^2_{\ell_1\ell_2\ell}\,b_{\ell_1}\,b_{\ell_2}\,b_{\ell}. 
\label{eq:odd3}
\een
The summations over possible modes defined by $(\ell,\ell_1,\ell_2)$ is restricted to $\ell+\ell_1+\ell_2={\rm odd}$ (o). Notice that the definition of reduced bispectrum given in Eq. (\ref{eq:bispec_def}) enforces   $\ell+\ell_1+\ell_2={\rm even}$ (e) i.e. includes only even-parity modes. These results can be generalized to the case of kurt-spectra. 

Clearly there is no obvious source that is expected to reach the signal-to-noise ratio of detectability.
However, such null-tests for odd-parity skew-spectra can definitely be included to check for possible
contamination from systematics, or as tests for as-yet unknown new physics.

The odd- \ah{and even-}parity {\em optimized} skew-spectra can likewise be expressed as:
\ben
&& S_\ell^{(\rm o)} = \sum_{\cb{\ell_1 \ge \ell_2}} {\hat B_{\ell_1\ell_2\ell}B_{\ell_1\ell_2\ell}\over \myC_{\ell_1}\myC_{\ell_2}\myC_{\ell}};
\quad \ell+\ell_1+\ell_2={\rm odd}; \quad\quad\\
&& S_\ell^{(\rm e)} = \sum_{\cb{\ell_1 \ge \ell_2}} {\hat B_{\ell_1\ell_2\ell}B_{\ell_1\ell_2\ell}\over \myC_{\ell_1}\myC_{\ell_2}\myC_{\ell}};
\quad \ell+\ell_1+\ell_2={\rm even}.
\een
In the first case only odd modes are included while in the second case we restrict 
to even modes, thus reducing it to the usual skew-spectrum described in \cite{MuHe10}.
If we further assume that the even and odd parity contributions can be separated with respective
amplitudes given by $f^{(\rm e)}_{\rm NL}$ and $f^{(\rm o)}_{\rm NL}$ for a specific model of non-Gaussianity,
we can write:
\ben
&& B_{\ell_1\ell_2\ell_3} = f^{\rm (e)}_{\rm NL}B^{\rm (e)}_{\ell_1\ell_2\ell_3} + 
f^{\rm (o)}_{\rm NL}B^{\rm (o)}_{\ell_1\ell_2\ell_3}.
\een
The estimators for $f^{(\rm e)}_{NL}$ and $f^{(\rm o)}_{NL}$ are given by:
\ben
&& f^{\rm (e)/(o)}_{NL} = {1 \over{\rm N}^{\rm (e)/(o)}}\sum_{\ell_1 \ge \ell_2 \ge \ell_3} {\hat B_{\ell_1\ell_2\ell_3}
B^{}_{\ell_1\ell_2\ell_3}\over \myC_{\ell_1}\myC_{\ell_2}\myC_{\ell_3}};
\quad \ell+\ell_1+\ell_2={\rm even/odd}; \quad\quad
{\rm N}^{\rm (e)/(o)} = \sum_{\ell_1 \ge \ell_2 \ge \ell_3} {\hat B_{\ell_1\ell_2\ell_3}B^{(e)/(o)}_{\ell_1\ell_2\ell_3}\over \myC_{\ell_1}\myC_{\ell_2}\myC_{\ell_3}}.
\een
The generalization to include partial sky coverage and to handle the inhomogeneous noise can be
done following the prescription in \cite{MuHe10}. 

\cb{Generalization to the case of odd-parity
kurt-spectra can be done in a similar manner. We start by noting that all-sky pairing function 
$P^{\ell_1\ell_2}_{\ell_3\ell_4}(\ell)$
and the flat-sky version $p$ can be linked by the following expression: $P^{\ell_1\ell_2}_{\ell_3\ell_4}(\ell) = I_{\ell_1\ell_2\ell}I_{\ell_3\ell_4\ell}\;\;p^{\ell_1\ell_2}_{\ell_3\ell_4}(\ell)$.
Using Eq. (\ref{eq:odd_bi}) we modify this expression as: $P^{\ell_1\ell_2}_{\ell_3\ell_4}(\ell) = {\cal I}_{\ell_1\ell_2\ell}{\cal I}_{\ell_3\ell_4\ell}\;\;p^{\ell_1\ell_2}_{\ell_3\ell_4}(\ell)$. Using this modified 
$P^{\ell_1\ell_2}_{\ell_3\ell_4}(\ell)$ and other terms that are obtained by permutations
of indices, e.g. $P^{\ell_1\ell_3}_{\ell_2\ell_4}(\ell)$ and$P^{\ell_1\ell_4}_{\ell_2\ell_3}(\ell)$  one can finally construct the
$[T^{(i)}]^{\ell_1\ell_2}_{\ell_3\ell_4}(\ell)$ that will match with the ordinary trispectrum
when $\Sigma_U=$ even and $\Sigma_L=$ even condition is satisfied but will not be vanishing
when one of these conditions is violated. The total trispectrum can be constructed from
four different contributions:
\ben
&& [T^{(i)}]^{\ell_1\ell_2}_{\ell_3\ell_4}(\ell)= \alpha\;[T^{(e-e),(i)}]^{\ell_1\ell_2}_{\ell_3\ell_4}(\ell)
+ \beta\;[T^{(e-o),(i)}]^{\ell_1\ell_2}_{\ell_3\ell_4}(\ell) + \gamma\;[T^{(o-e),(i)}]^{\ell_1\ell_2}_{\ell_3\ell_4}(\ell)
+ \delta\;[T^{(o-o),(i)}]^{\ell_1\ell_2}_{\ell_3\ell_4}(\ell).
\een
Here, ($\alpha,\beta,\gamma,\delta$) define the relative contributions from four different
types of trispectra. We can now define the odd-parity kurt-spectra:
\ben
&& K_{\ell}^{(i)} = \sum_{\ell_1 \ge \ell_2 \ge \ell_3 \ge \ell_4}{1 \over \Xi_{\ell}} [T^{(i)}]^{\ell_1\ell_2}_{\ell_3\ell_4}(\ell)
{\cal J}_{\ell_1\ell_2\ell}{\cal J}_{\ell_3\ell_4\ell}; 
\quad\quad {\cal J}_{\ell_1\ell_2\ell} =  {{\cal I}_{\ell_1\ell_2\ell}\over \Xi_{\ell}}.
\een
Depending on whether $\Sigma_U=\ell_1+\ell_2+\ell$ and $\Sigma_L=\ell_3+\ell_4+\ell$ are 
restricted to even(e) or odd(o) we have 
four different possible combinations i.e. $[K_{\ell}^{(i)}]^{(\rm o-o)}$ when $\Sigma_U$ and $\Sigma_L$ are both
odd and similarly one can have $[K_{\ell}^{(i)}]^{\rm (o-e)}$ and $[K_{\ell}^{(i)}]^{\rm (e-o)}$ or
$[K_{\ell}^{(i)}]^{\rm (e-e)}$ for other possible choices. 
The estimator $[K_{\ell}^{(i)}]^{\rm (e-e)}$ denotes the usual choice in the literature.
The modifications of the optimised kurt-spectra defined in \cite{Mu_kurt10} can be done using the same 
techniques e.g. using suitably-defined optimized kurt-spectra associated with the odd-odd parity trispectra we have the following estimators for the amplitude $\delta$:
\ben
&& \delta = {1 \over N^{(o-o)}} \sum_{\ell_1\ge \ell_2 \ge \ell_3 \ge \ell_4} {1 \over \Xi_{\ell}}{ 
\hat T_{\ell_3\ell_4}^{\ell_1\ell_2}(\ell)T_{\ell_3\ell_4}^{\ell_1\ell_2}(\ell) 
\over \myC_{\ell_1}\myC_{\ell_2}\myC_{\ell_3}\myC_{\ell_4}}; \quad 
\Sigma_U = {\rm odd}, \Sigma_L = {\rm odd}; \quad\quad
N^{\delta} =  \sum_{\ell_1\ge \ell_2 \ge \ell_3 \ge \ell_4}{1 \over \Xi_{\ell}}{ 
\hat T_{\ell_3\ell_4}^{\ell_1\ell_2}(\ell)T_{\ell_3\ell_4}^{\ell_1\ell_2}(\ell) 
\over \myC_{\ell_1}\myC_{\ell_2}\myC_{\ell_3}\myC_{\ell_4}}.
\een
In the above expression we have restricted both the triplets defined by $\Sigma_U$ and $\Sigma_L$
to odd-parity modes. The normalisation $N^{\delta}$ is also defined using the same restrictions.
The estimators for $\alpha$, $\beta$ and $\gamma$  can also be constructed in an analogous manner.} 

\section{Minkowski Functionals in a Needlet basis }
\label{sec:needlet}
The use of wavelets in the CMB is now well established \citep{FS98,AV99,Mc06,Mc07}. 
Wavelet analysis provides an intermediate choice 
between real-space analysis and analysis in the harmonic domain and is particularly suitable
for localised signals. Needlets are special types of spherical wavelets that allow localised filtering in both
real space and in the harmonic domain. They have compact support in the harmonic domain but are still very
well localised in the pixel basis \citep{NPW06,Mar08,GFC07}. It has been used previously for foreground subtraction \citep{Han06}, 
component separation \citep{St06,BJ12}, point-source detection \citep{Sanz06} polarisation analysis \citep{Cab07} 
as well as testing non-Gaussianity \citep{Vielva04,Cab04, Rud09,simona12} and detection of features in the CMB sky
\citep{Pier08}.
We start with the decomposition of a generic function $\Psi(\oh)$ using a needlet basis $\Phi_{jk}(\oh)$
\ben
\Psi(\oh) = \sum_{jk} \Psi_{jk}\, \Phi_{jk}(\oh); \quad\quad \Phi_{jk}(\oh) = \sqrt{\lambda_{jk}}
\sum_{\ell} 
\varpi_{\ell}^{(j)}\sum_{m=-\ell}^{\ell}Y^*_{\ell m}(\oh) Y_{\ell m}(\hat\Omega_{jk}); \quad\quad
 \varpi^{(j)}_{\ell} = \varpi\left ( {\ell \over {\cal B}^j} \right ).
\label{eq:needlet_transform}
\een
Here $\{\oh_{jk}\}$ defines a set of {\em cubature points} on the unit sphere corresponding to frequency $j$,
and $\{\lambda_{jk}\}$ denotes the {\em cubature weights}. The needlet coefficients  $\{\lambda_{jk}\}$ are 
proportional to the pixel area. For a given HEALPix\footnote{http://healpix.jpl.nasa.gov} resolution the centres of pixels can serve as curbature points and $\lambda_{jk} = ({4\pi / N^j_{\rm pix}})$ where $N^j_{\rm pix}$ is the total number
of pixels at a given resolution. The {\em weight function or filter} $\varpi(t)$ satisfies three different conditions. {\bf Compact support:} $\varpi(t)$ is strictly positive in the interval 
$t \in [{\cal B}^{-1},{\cal B}]$ for a given ``dilation parameter'' ${\cal B}$, thus $\varpi_{\ell}^{(j)}$ has support in 
$\ell \in [{\cal B}^{j-1},{\cal B}^{j+1}]$. {\bf Smoothness:} $\varpi(t)$ is infinitely differentiable in (0,$\infty$), and finally
{\bf partition of unity:} for any given  $\ell$, we have 
$\sum_j [\varpi^{(j)}_\ell]^2 = 1$ for all $\ell > {\cal B}$. The specific recipe for constructing $\varpi_{\ell}^{(j)}$ can be found in \cite{Mar08}.
The needlet coefficients $\Psi_{jk}$ are given by the inverse needlet transform
and are expressed in terms of the harmonic coefficents $\Psi_{\ell m}$ of the map $\Psi(\oh)$:
\ben
\Psi_{jk} \equiv \;\int d\oh\; \Phi_{jk}(\oh)\Psi(\oh) = \sqrt{\lambda_{jk}}\;\sum_\ell \varpi_{\ell}^{(j)}\sum_{m=-\ell}^{\ell}
\Psi_{\ell m}b_{\ell} \; Y_{\ell m}(\hat\Omega_{jk}).
\een
The above expression relates the needlet coefficients $\Psi_{jk}$ to the harmonic coefficients
$\Psi_{\ell m}$. The power-spectrum  $\myC_{\ell}$ and the {\em needlet power-spectrum} $\beta^{(1,1)}_j$ are related through the
following expression:
\ben
\beta_j^{(\Psi,\Psi)} \equiv {{1 \over N_{\rm pix}^{(j)}}}\sum_k\Psi_{jk}\Psi^*_{jk} = 
\sum_{\ell}[ \varpi^{(j)}_{\ell} ]^2 {\Xi_{\ell} \over 4 \pi}\myC_{\ell}b_l^2; \quad\quad \sum_j \beta_j^{(\Psi,\Psi)}
= \sum_{\ell} {\Xi_{\ell} \over 4 \pi}\myC_{\ell}b^2_{\ell} = \la [\delta \Psi(\oh)]^2\ra.
\een
Thus the needlet power-spectrum is simply the variance computed using a specific set of {\em filters} $\varpi^{(\ell)}_j$.
The needlet power spectrum computed with partial sky coverage can likewise be expressed in terms of the 
harmonic power-spectrum obtained from the partial sky coverage:
\ben
\tilde \beta_j^{(\Psi,\Psi)} \equiv {{1 \over N_{\rm pix}^{\ah{(j)}}}\sum_k\ah{\tilde \Psi_{jk}\tilde \Psi^*_{jk}} = 
\sum_{\ell}[ \varpi^{(j)}_{\ell} ]^2 {\Xi_{\ell} \over 4 \pi}\tilde \myC_{\ell}b^2_{\ell}}.
\een
The convolved power spectrum recovered from the partial sky $\tilde \myC_{\ell}$ can be
expressed in terms of the all-sky power spectrum $\tilde \myC_{\ell} = {\mathbb M}_{\ell\ell'}\myC_{\ell'}$ 
(${\mathbb M}_{\ell\ell'}$ is defined in Eq. (\ref{eq:alm_est})). It is possible to express the
needlet power spectrum $\tilde\beta_j$ from the partial sky in terms of the all-sky power spectra
 $\tilde\beta_j$ which allows definition of an unbiased estimator.
\ben
\tilde \beta_j = \sum_{j'} {\mathbb T}_{jj'} \beta_j; \quad\quad 
\hat \beta_j = \sum_{j'}{\mathbb T}^{-1}_{jj'}\tilde \beta_{j'}; 
\een
\ah{where we define}
\ben
{\mathbb T}_{jj'} = \sum_{\ell'}{\mathbb K}_{j\ell'} {\mathbb S}^{-1}_{\ell'j'}; \quad\quad
{\mathbb K}_{\ell' j} = \sum_{\ell} [\varpi^{(j)}_{\ell}]^2(2\ell+1){\mathbb M}_{\ell\ell'}; \quad\quad
{\mathbb S}_{\ell j} = (2\ell+1)[\varpi_{\ell}^{(j)}]^2.
\een
A similar construction $\beta_j^{(\Psi,\Psi')}$ involving two different fields $\Psi$ and $\Psi'$  
is possible which will depend on the 
cross-spectra involving these two fields $\la\Psi_{\ell m}\Psi'_{\ell m}\ra = \myC^{\Psi\Psi'}_{\ell}\delta_{\ell\ell'}\delta_{mm'}$. Such 
{\em needlet cross-spectra} have already been used in cross-correlating large-scale tracers such as the maps of galaxy 
distributions from the surveys such as NVSS and CMB maps from WMAP to study the Integrated Sachs-Wolfe effect \citep{PBM06}.

The {\em needlet bispectrum} $S_{j_1j_2j_3}$ and {\em trispectrum} can similarly be expressed in terms of the bispectrum $B_{\ell_1\ell_2\ell_3}$ and trispectrum using the following expressions:
\ben
&& S_{j_1j_2j_3} = {1 \over N^{(j)}_{\rm pix}}\sum_k \Psi_{j_1k}\Psi_{j_2k}\Psi_{j_3k}
=\sum_{\ell_1\ell_2\ell_3}\varpi^{(j_1)}_{\ell_1}\varpi^{(j_2)}_{\ell_2}\varpi^{(j_3)}_{\ell_3} I_{\ell_1\ell_2\ell_3}B_{\ell_1\ell_2\ell_3}
b_{\ell_1}b_{\ell_2}b_{\ell_3}; \\ 
&& T_{j_1j_2j_3j_4} = {1 \over N^{(j)}_{\rm pix}}\sum_k \Psi_{j_1k}\Psi_{j_2k}\Psi_{j_3k}\Psi_{j_4k}
=\sum_{\ell_1\ell_2\ell_3}\varpi^{(j_1)}_{\ell_1}\varpi^{(j_2)}_{\ell_2}\varpi^{(j_3)}_{\ell_3}
\varpi^{(j_4)}_{\ell_4} \sum_{\ell}\; {1\over\Xi_\ell}\; I_{\ell_1\ell_2\ell}I_{\ell_3\ell_4\ell}
T^{\ell_1\ell_2}_{\ell_3\ell_4}(\ell)b_{\ell_1}b_{\ell_2}b_{\ell_3}b_{\ell_4}.
\een
Thus the needlet bispectrum or trispectrum is equivalent to the ordinary skewness or kurtosis with 
varying weights specified by the indices $\{j_{i}\}$. Note that individual 
estimates of the needlet bispectrum $S_{j_1j_2j_3}$ are expected to be noise-dominated. 

Next, we introduce here the concept of the skew-spectra in the needlet domain. We expand the maps $\Psi^2, \nabla^2\Psi$
and $\nabla\Psi\cdot\nabla\Psi$  in their needlet basis:
\ben
&& [\Psi^2]_{jk} = \int d\oh \; \Phi_{jk}(\oh) [\Psi^2(\oh)]
 = \int d\oh \; \Phi_{jk}(\oh)\;
\sum_{j_1k_1}\Psi_{j_1k_1}\Phi_{j_1k_1}(\oh)\sum_{j_2k_2}\Psi_{j_2k_2}\Phi_{j_2k_2}(\oh)\nn \\
&& \quad\quad\quad =\sum_{j_1k_1}\sum_{j_2k_2}\Psi_{j_1k_1}\Psi_{j_2k_2}\sqrt{\lambda_{jk}}\sqrt{\lambda_{j_1k_1}}\sqrt{\lambda_{j_2k_2}}
\sum_{\ell_1\ell_2}\varpi^{(j)}_{\ell}\varpi^{(j_1)}_{\ell_1}\varpi^{(j_2)}_{\ell_2} I_{\ell_1\ell_2\ell} \;\nn \\
&& \quad\quad\quad\times \sum_{mm_1m_2}\left ( \begin{array}{ c c c }
     \ell_1 & \ell_2 & \ell \\
     m_1 & m_2 & m
  \end{array} \right)
Y_{\ell_1 m_1}(\oh_{j_1k_1})Y_{\ell_2 m_2}(\oh_{j_2k_2})Y_{\ell m}(\oh_{jk}); \\
&& {[\nabla^2\Phi]}_{jk}= \int d\oh \; \Phi_{jk}(\oh) [\nabla^2\Psi(\oh)]
= -\sqrt{\lambda_{jk}} \sum_\ell \varpi^{(j)}_{\ell} \Pi_{\ell} b_\ell
\sum_{m=-\ell}^{\ell}\Psi_{\ell m}Y_{\ell m}(\oh_{jk});\nn \\
&& {[\nabla \Psi \cdot \nabla\Psi]}_{jk} =\int d\oh \;\Phi_{jk}(\oh) [\nabla\Psi(\oh)\cdot\nabla\Psi(\oh)]= \int d\oh \; \Phi_{jk}(\oh) \sum_{j_1k_1}\Psi_{j_1k_1}\nabla\Phi_{j_1k_1}(\oh)\cdot\sum_{j_2k_2}\Psi_{j_2k_2}\nabla\Phi_{j_2k_2}(\oh)\\
&& \quad\quad\quad ={1 \over 3}\sum_{j_1k_1}\sum_{j_2k_2}\Psi_{j_1k_1}\Psi_{j_2k_2}\sqrt{\lambda_{jk}}\sqrt{\lambda_{j_1k_1}}\sqrt{\lambda_{j_2k_2}}
\sum_{\ell_1\ell_2}\varpi^{(j)}_{\ell}\varpi^{(j_1)}_{\ell_1}\varpi^{(j_2)}_{\ell_2} 
(\Pi_{\ell_1}+\Pi_{\ell_2}-\Pi_{\ell}) I_{\ell_1\ell_2\ell} \;\nn \\
&& \quad\quad\quad\times \sum_{mm_1m_2}(-1)^m\left ( \begin{array}{ c c c }
     \ell_1 & \ell_2 & \ell \\
     m_1 & m_2 & m
  \end{array} \right)
Y_{\ell_1 m_1}(\oh_{j_1k_1})Y_{\ell_2 m_2}(\oh_{j_2k_2})Y_{\ell m}(\oh_{jk}). 
\een
We now define the skew-spectra, labelled by $j$, in the needlet basis by the following expressions:
\ben
&& \beta_j^{(\Psi^2,\Psi)} = {1 \over {N^{(j)}_{\rm pix}}}\sum_k{ [\Psi^2]}_{jk}\Psi_{jk} 
= \sum_{\ell\ell_1\ell_2} 
[\varpi_{\ell}^{(j)}]^2 \; I_{\ell_1\ell_2\ell} B_{\ell_1\ell_2\ell}\; b_{\ell_1}b_{\ell_2}b_{\ell} \;;
\label{eq:skew_spec_need1}\\
&& \beta_j^{(\Psi^2,\nabla^2\Psi)} = {1 \over {N^{(j)}_{\rm pix}}}\sum_k {[\Psi^2]}_{jk}[\nabla^2\Psi]_{jk} 
= \sum_{\ell\ell_1\ell_2} 
[\varpi_{\ell}^{(j)}]^2 \; (\Pi_{\ell_1}+\Pi_{\ell_2}+\Pi_{\ell})I_{\ell_1\ell_2\ell} B_{\ell_1\ell_2\ell} \; b_{\ell_1}b_{\ell_2}b_{\ell}\;;
\label{eq:skew_spec_need2}\\
&& \beta_j^{(\nabla^2\Psi,\nabla\Psi\cdot\nabla\Psi)} = {1 \over {N^{(j)}_{\rm pix}}}
\sum_k {[\nabla\Psi\cdot\nabla\Psi]}_{jk} [\nabla^2\Psi]_{jk} 
=\sum_{\ell\ell_1\ell_2} 
[\varpi_{\ell}^{(j)}]^2 \;(\Pi_{\ell} + \Pi_{\ell_1} -\Pi_{\ell_2})\Pi_{\ell_2} + {\rm cyc.perm.}) I_{\ell_1\ell_2\ell} B_{\ell_1\ell_2\ell} \; b_{\ell_1}b_{\ell_2}b_{\ell}\;.
\label{eq:skew_spec_need3}
\een
Thus the three generalised skew-spectra can be obtained simply by cross-correlating the relevant maps in the needlet basis. 
No specific assumption about the underlying bispectrum is used. Though the primary aim is to study the primordial
bispectrum, the results will be equally applicable to that due to the secondaries. The main advantage of using
the skew-spectra is related to its ability to probe the non-Gaussianity as a function of angualr scale thus
retaining the power to discriminate between various models of primaries or secondaries.

Using Eq. (\ref{eq:bispec}) the coefficients of modal decomposition recovered from maps can also be used to estimate directly
the skew-spectra in the needlet domain, and results from two different methods can be useful for consistency checks.

In our derivation we have used the following expression for the needlet bispectrum:
\ben
\la \Psi_{j_1k_{\ah{1}}}\Psi_{j_2k_2}\Psi_{j_3k_3}\ra &=& \sqrt{\lambda_{j_1k_1}}\sqrt{\lambda_{j_2k_2}}\sqrt{\lambda_{j_3k_3}}
\sum_{\ell_1\ell_2\ell_3} \varpi^{(j_1)}_{\ell_1}\varpi^{(j_2)}_{\ell_2}\varpi^{(j_3)}_{\ell_3} \nn \\
&&\times \sum_{m_1m_2m_3}Y_{\ell_1m_1}(\oh_{j_1k_1})Y_{\ell_2m_2}(\oh_{j_2k_2})Y_{\ell_3m_3}(\oh_{j_3k_3})
\left ( \begin{array}{ c c c }
     \ell_1 & \ell_2 & \ell_3 \\
     m_1 & m_2 & m_3
  \end{array} \right)B_{\ell_1\ell_2\ell_3} b_{\ell_1}b_{\ell_2}b_{\ell_3}.
\een

Finally we can express the needlet skew-spectra
as a convolution of the skew-spectra defined in the harmonic domain.
\ben
\beta_j^{(\Psi^2,\Psi)} =  \sum_{\ell} \Xi_{\ell} \; [\varpi_{\ell}^{(j)}]^2 \; S^{(0)}_{\ell}; \quad\quad
\sum_j \beta_j^{(\Psi^2,\Psi)} = \sum_{\ell} \; \Xi_{\ell} \; S^{(0)}_{\ell} = S^{(0)}.
\label{eq:need_skew}
\een
Similar results can be obtained relating $\beta_j^{(\Psi^2,\nabla^2\Psi)}$ to  $S^{(1)}_{\ell}$
as well as $\beta_j^{(\nabla\Psi\cdot\nabla\Psi,\nabla^2\Psi)}$ to 
$S^{(2)}_{\ell}$.
Thus the needlet skew-spectra are simply {\em binned harmonic skew-spectra}.
\begin{figure}
\begin{center}
{\epsfxsize=10. cm \epsfysize=5.0 cm 
{\epsfbox[30 429 583 715]{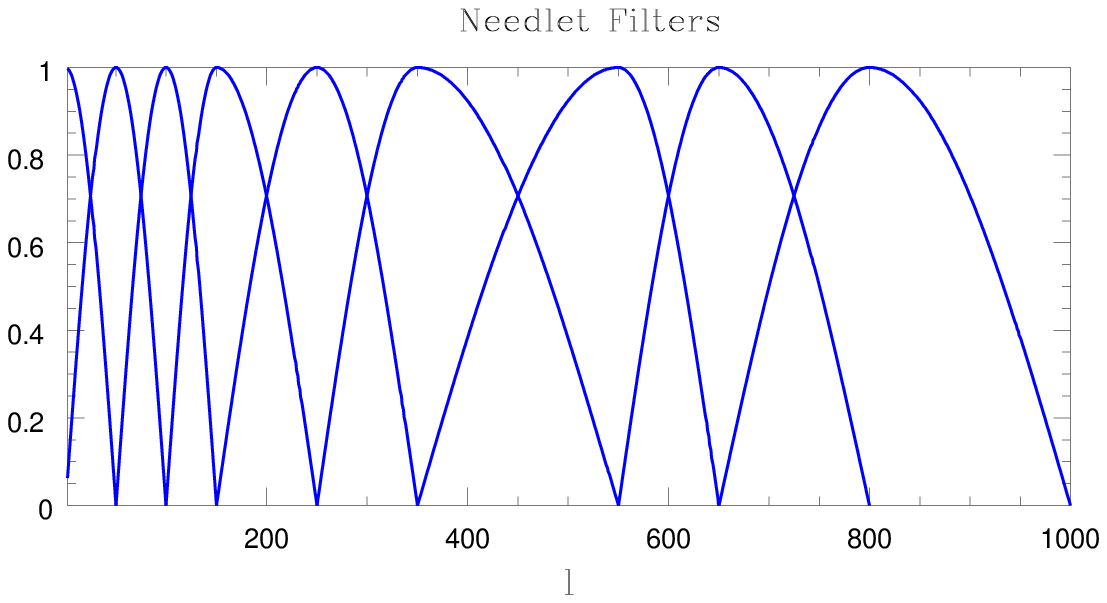}}}
\end{center}
\caption{The needlet filters $\varpi_\ell^{(j)}$ defined in Eq. (\ref{eq:needlet_transform}) that correspond to nine different {\em frequencies} $j$ used in our study are depicted.}
\label{fig:need_beam}
\end{figure}
The higher-order generalization to two-to-two and three-to-one power-spectra 
can be achieved using the same principle:
\ben
&& \beta_j^{(\Psi^2,\Psi^2)} = {1 \over {N^{(j)}_{\rm pix}}}\sum_k [\Psi^{2}]_{jk}[\Psi^{2}]_{jk} 
= \sum_{\ell}\Xi_{\ell}[\varpi_{\ell}^{(j)}]^2 K_\ell^{(2,2)}  =\sum_{\ell} [\varpi_{\ell}^{(j)}]^2
\; \sum_{\ell_{i}} {1 \over \Xi_L} I_{\ell_1\ell_2\ell} I_{\ell_3\ell_3\ell}T^{\ell_1\ell_2}_{\ell_3\ell_4}(\ell)b_{\ell_1}b_{\ell_2}b_{\ell_3}b_{\ell_4}; 
\label{eq:need22}\\
&& \beta_j^{(\Psi^3,\Psi)} =  {1 \over {N^{(j)}_{\rm pix}}}\sum_k [\Psi^{3}]_{jk}[\Psi^{}]_{jk}= \sum_{\ell} \Xi_{\ell} [\varpi_{\ell}^{(j)}]^2 K_{\ell}^{(3,1)}
\; = \sum_{\ell} \;\Xi_{\ell}\; [\varpi_{\ell}^{(j)}]^2\sum_{\ell_{i}\ell} {1 \over \Xi_\ell} 
I_{\ell_1\ell_2\ell} I_{L\ell_3\ell}\; T^{\ell_1\ell_2}_{\ell_3\ell}(\ell)
b_{\ell_1}b_{\ell_2}b_{\ell_3}b_{\ell}.
\label{eq:need31}
\een
In both cases if we sum over all possible modes we can recover the kurtosis $K_4$:
\ben
\sum_j \beta_j^{\Psi^2,\Psi^2} =\sum_j \beta_j^{\Psi^3,\Psi} =  K_4 
 = \sum_{\ell_i}\sum_{\ell}  I_{\ell_1\ell_2 L} I_{\ell_3\ell L}T^{\ell_1\ell_2}_{\ell_3\ell_4}(\ell)
b_{\ell_1}b_{\ell_2}b_{\ell_3}b_{\ell_4}. 
\label{eq:needk4}
\een
Thus we arrive at the same one-point kurtosis using a different modal expansion. The
generalised kurt-spectra required to construct the MFs have a similar expression:
\ben
\beta_j^{(i)} = \sum_{\ell} \Xi_{\ell}\; [\varpi_{\ell}^{(j)}]^2 \; K_{\ell}^{(i)};
\quad\quad \sum_j \beta_j^{(i)} = K^{(i)}.
\label{eq:need_kurt}
\een
The expression that relates $K_{\ell}^{(i)}$ with the corresponding generalised
trispectrum is given in Eq. (\ref{eq:kurt_spectra2}) . The expressions for the generalised trispectra are 
given in Eq. (\ref{eq:kurt_first})-Eq. (\ref{eq:kurt_spectra1}).

To perform an error analysis we note that the error in an arbitrary needlet 
skew- or kurt-spectrum $\beta_j^{A,B}$ can be expressed as a weighted sum of 
scatter in respective skew- or kurt-spectra $S^{(A,B)}_{\ell}$ or $K^{(A,B)}_{\ell}$:
\ben
\delta \beta^{(A,B)}_j = \sum_{\ell} \Xi_{\ell}[\varpi_{\ell}^{(j)}]^2\delta S^{(A,B)}_{\ell}.
\label{eq:need_error}
\een
Using expressions of error-covariance for $\delta S^{(A,B)}_{\ell}$ or $\delta T^{(A,B)}_{\ell}$ derived earlier, we can
work out similar expressions for the needlet spectra $\delta \beta^{(A,B)}_j$.
\ben
\mathbb{C}^{(A,B)}_{ij} = \sum_{\ell}\sum_{\ell'} \Xi_{\ell}[\varpi_{\ell}^{(i)}]^2 \mathfrak{C}^{(A,B)}_{\ell\ell'} \Xi_{\ell'}[\varpi_{\ell'}^{(j)}]^2,
\een
with $\mathbb{C}^{(A,B)}_{ij}=\la \delta \beta^{(A,B)}_i \beta^{(A,B)}_j \ra$ and 
$\mathfrak{C}^{(A,B)}_{\ell\ell'}= \la\delta \myC^{(A,B)}_l \delta\myC^{(A,B)}_{l'}\ra$ being the covariance in needlet and
harmonic bases.

A few comments are in order at this point. The cumulant correlators and the series of multi-spectra such as skew- or kurt spectra 
represent two-point objects in real space and in the harmonic domain. The statistics 
developed here are their needlet representation. Throughout the results here ignore odd {\em parity} modes.
To include the {\em parity odd} modes in Eq. (\ref{eq:skew_spec_need1})-Eq. (\ref{eq:skew_spec_need3}) 
we can  replace $I_{\ell_1\ell_2\ell_3}$ with ${\cal I}_{\ell_1\ell_2\ell_3}$ and use 
$B_{\ell_1\ell_2\ell_3} = b_{\ell_1\ell_2\ell_3}{\cal I}_{\ell_1\ell_2\ell_3}$. 
The discussion in this section does not depend on a specific form of the bispectrum and can be relevant in many other areas in cosmology, e.g.
weak lensing or studies regarding galaxy distribution where studies of non-Gaussianity is performed.

The concept of needlets has been extended by \cite{GM09a,GM09b} to Mexican needlets by replacing the compactly-supported 
kernel $\varpi_{\ell}^{(j)}$ with a smooth functional form. Mexican needlets have extremely good
localization properties in real space and can be used to approximate the Spherical Mexican Wavelet at high
angular frequencies. Results derived above can be applicable in such situations.
\begin{figure}
\begin{center}
{\epsfxsize=16.5 cm \epsfysize=6.0 cm {\epsfbox[29 535 583 725]
{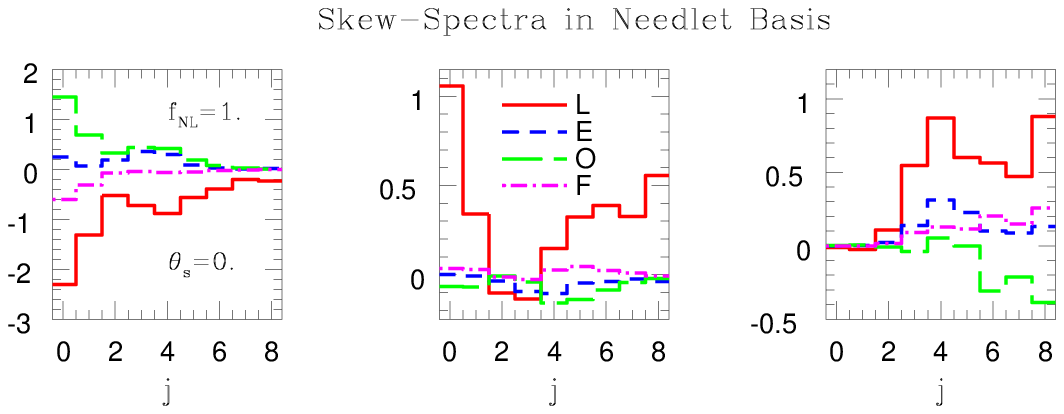 }}}
\caption{The three different skew-spectra defined in the needlet domain, 
by Eq. (\ref{eq:need_skew}) are shown for the bins defined in Fig-\ref{fig:need_beam}.
From left to right the different panels correspond to $S_j^{(0)}$, $S_j^{(1)}$ and $S_j^{(2)}$. 
The lines correspond to the local (L), equilateral (E), orthogonal (O) and folded (F)  models of primordial \cb{non-Gaussianity}. 
The experimental set up is that of the Planck 143 GHz channel and normalisation $f_{\rm NL}$ for each model is taken to be unity.}
\label{fig:need1_ps}
\end{center}
\label{fig:need2_ps}
\end{figure}
It was noticed previously that the needlet-based estimators are generally less affected by 
anisotropic noise and observational mask \citep{Curt11}. Further tightening of error-bars of a 
of a KSW-based optimized estimator and a linear correction term was reported by \cite{simona12}
which can be adopted in our analysis. The generalisation to the case of spinorial fields  by using a spin-needlet decomposition \citep{GM10}
will be presented elsewhere. 

For specific computation of skew-spectra in needlet basis we choose the following functional form \cite{BJ12}:
%
\begin{table*}
\begin{center}
\begin{tabular}{|c |c|c| c| c | c |c|c| c| c}
\hline
$j$ &0&1&2&3&4&5&6&7&8\\
\hline
$\ell_{\rm min}$ &0 & 0 & 50 & 100 & 150 & 250 & 350 & 550 & 650 \\
\hline
$\ell_{\rm max}$ & 50 & 100 & 150 & 250 & 350 & 550 & 650 & 800 & 1000\\
\hline
$\ell_{\rm peak}$ & 0 & 50& 100 & 150& 250 & 350 & 550 & 650 & 800\\
\hline
\end{tabular}
\caption{Specification of the filter functions $\varpi_\ell^{(j)}$ used. The functional form is given in Eq. (\ref{eq:filter}).}
\label{tab:need}
\end{center}
\end{table*}
\ben
&& \varpi^j_{\ell}=\sin\left[ {\ell_{\rm peak} -\ell \over \ell_{\rm peak} -\ell_{\rm min}} {\pi \over 2}\right]; \quad\quad \ell<\ell_{\rm peak};\nn\\
&& \varpi^j_{\ell}=1; \quad\quad  \ell=\ell_{peak};\nn\\
&& \varpi^j_{\ell}=\sin\left[ {\ell - \ell_{\rm peak} \over \ell_{\rm max} -\ell_{\rm peak}} {\pi \over 2}\right]; \quad\quad \ell>\ell_{\rm peak}. \label{eq:filter}
\een
The choice of $\ell_{\rm min}$, $\ell_{\rm max}$ and $\ell_{\rm peak}$ are tabulated in Table-\ref{tab:need} for nine different {\em frequencies}
which are plotted in Figure-\ref{fig:need_beam}. The resulting expressions for the skew-spectra in the needlet basis 
$\beta_j^{\psi^2,\Psi}, \beta_j^{\Psi^2,\nabla^2\Psi}$ and $\beta_j^{\nabla\Psi\cdot\nabla\Psi,\nabla^2\Psi}$ are shown in Figure-\ref{fig:need1_ps} for {\em local} and {\em equilateral} models of primordial non-Gaussianity.
\section{Discussion of Results}
\label{sec:disc}
Different approaches are essential while testing non-Gaussianity
as they all exploit different statistical characteristics. There
is no unique approach that can be adopted to describe or parametrize
non-Gaussianity in a complete manner. Testing of non-Gaussianity 
therefore must be done using a battery of complementary techniques, and each 
of these techniques has a unique response to the real-world
issues such as the sky-coverage and instrumental noise.
Any robust detection therefore will have to involve a
simultaneous cross-validation of results obtain from independent methods.

{\bf 1. Generalised Skew Spectra to differentiate different Primordial or Secondary Contribution:} 
One of the main difficulties faced by one-point estimators, which 
also include\cb{s} the MF-based approaches, is their inability
to differentiate various sources of non-Gaussianity. The
one-point estimators typically compress all available 
information to a single number.   \ah{If different sources of non-Gaussianities are considered simultaneously, then the compression is typically to a set of numbers equal to the number of parameters to be estimated (see e.g. \cite{Komatsu09}), but in this case verification that the non-gaussian sources probed are actually responsible for the non-Gaussian signals is not possible.} 
This has the advantage of increasing the signal-to-noise ratio but it loses the
ability to differentiate various sources of non-Gaussianity.
In any study of primordial non-Gaussianity it is of the
utmost importance to avoid any cross-contamination from
secondary sources (see e.g. \cite{GoldbergSpergel99a,GoldbergSpergel99b, CoorayHu}). In recent studies involving MFs,
a certain level of disagreement has been noticed with studies
that use the bispectrum to probe non-Gaussianity (see e.g. \cite{HKM06}).
Given that MFs-based approaches only directly probe the bispectra, as the contributions from the higher-order multi-spectra
are subdominant, it is important to understand the
reasons for these disagreements.

Following \cite{MuHe10}, we have developed a new technique to study the morphology of the
CMB sky. Instead of one-point estimators, e.g. the skewness,
their method relies on a spectrum or skew-spectrum which is the Fourier
transform of two-point objects in real space known as cumulant correlators. These 
skew-spectra do not compress all available information
from the study of a bi-spectrum to a single number and their shape 
can help to distinguish among various sources of non-Gaussianity.
Exploiting the perturbative expansion of the MFs it can \cb{be} seen 
that at the leading order of non-Gaussianity, the
MFs depend on three generalized skewness parameters Eq. (\ref{eq:skewness_def}). We extend the 
concept of the skew-spectra to the study of these MFs and 
introduce one generalized skew-spectrum with each of these skewness parameters.
This allows us to introduce a  power spectrum
associated with MFs. The advantage of cross-checking the
contributions to MFs using the skew-spectra is that they provide a
method to test any contamination from  secondaries or foregrounds.
The methods based on the skew-spectra are simpler to implement once the derivative
maps are constructed. These methods are similar
to moment-based approaches for studying non-Gaussianity 
and hence can provide a valuable basis for cross-comparison.
We have shown that this can be implemented in a model-independent
way. Our method is based on a pseudo-$\myC_\ell$ approach and
can handle arbitrary sky coverage and inhomogeneous noise distribution.
The pseudo-$\myC_{\ell}$ approach is well understood in the context of 
power spectrum studies and its variance or scatter can be 
computed analytically. We provide generic analytical results for 
computation of scatter around the individual estimates in Eq. (\ref{eq:scat1})-Eq. (\ref{eq:scat3}).
The level of cross-correlation among various estimators can be estimated using  
Eq. (\ref{eq:cross1})-Eq. (\ref{eq:cross3}).

It is also possible to go beyond the lowest level in non-Gaussianity.
However, it is expected that such correction will be subdominant
at least in the context of CMB data analysis. Nevertheless 
we include the spectrum associated with the next-order correction terms that were 
introduced by \cite{Mat10}. These terms represent kurtosis and 
the corresponding spectra are known as kurt-spectra. In
their study \citep{Mu_kurt10} introduced two sets of kurt-spectra.
Adopting their method we show that generic two-to-two
kurt-spectra can be extracted from the data without introducing
any additional complication using Eq. (\ref{eq:direct1})-Eq. (\ref{eq:kurt_spectra}). 
Using a simple model for the bispectrum from unsubtracted points sources Eq. (\ref{eq:point})
it is possible to provide an estimate of cross-contamination from this foreground
in estimation of other types of non-Gaussianity.

The one-point generalised skewness parameters $S^{(i)}$ are plotted in Figure \ref{fig:skewness_onept}.
In Figure \ref{fig:S_loc} we have shown the result of our computation for generalized skew-spectra
$S^{(0)}_{\ell}$, $S^{(1)}_{\ell}$, $S^{(2)}_{\ell}$ for various beam sizes $\theta_s$. 
In Figure \ref{fig:S_eq} the results
for equilateral model are shown, with the corresponding normalization set to unity $f^{\rm equi}_{\rm NL}=1$
and Figure \ref{fig:S_or} and Figure \ref{fig:S_en} correspond to  orthogonal and enfolded models respectively.
The corresponding results for point sources are depicted in Figure \ref{fig:S_ps}.
In Figure \ref{fig:cumu_loc} we show the three cumulant correlators corresponding to the Planck experiment.
The cumulant correlators are real-space representations of the corresponding skew-spectra.
We have plotted the S/N for various primordial models of non-Gaussianity in Fig. \ref{fig:scatter_loc}-\ref{fig:scatter_orth} for various non-Gaussian models.
%

{\bf 2. Generalised skew-spectra in the needlet basis:} 
In addition to their usual harmonic-domain representation and characterization in real space
using the cumulant correlators, we have analyzed the skew-spectra in the the needlet domain. The skew-spectra defined in
needlet domain are intermediate between the cumulant correlators defined in real
space and the skew-spectra defined in harmonic domain. It allows localized filtering
in real as well as in harmonic domain.  In Appendix-\ref{sec:needlet} we show that the skew-spectra in the needlet
domain can be obtained through appropriate filtering of the skew-spectra in the harmonic
domain Eq. (\ref{eq:skew_spec_need1})-Eq. (\ref{eq:skew_spec_need3}). These expressions can be 
used to construct the MFs in  the needlet domain using the expressions for the
skewness parameters Eq. (\ref{eq:need_skew}). Though mathematically equivalent, use of different 
bases can be useful for understanding the impact of various systematics.

We also relate both of the kurt-spectra, two-to-\ah{two} and three-to-one defined in the needlet basis in terms of their counterparts
in the harmonic domain in Eq. (\ref{eq:need22}) and Eq. (\ref{eq:need31}), which can both be used to
construct the kurtosis using Eq. (\ref{eq:needk4}). The generalized kurt-spectra can also be
constructed using similar techniques and are filtered versions of their harmonic counterparts.

Errors in the needlet basis can be related to their harmonic counterparts via Eq.(\ref{eq:scat3}).
Estimation of MFs can be done using the HEALPix-based pipeline that uses 
publicly-available software such as NeedaTool \citep{Pier10}. For the estimation of needlet
skew- and kurt-spectrum we provide a pseudo-$\myC_{\ell}$ based approach.

The needlet filters $\varpi_\ell^{(j)}$ used in our study are presented in Figure \ref{fig:need_beam}. 
and the needlet skew-spectra $S_j^{(0)},S_j^{(1)}$ and
$S_j^{(2)}$ for Planck- and EPIC-type experiments are given in Figure C1 and Figure C2 respectively.

{\bf 3 Generalised skew-spectra and modal decomposition:}
We have used the coefficients from the modal decomposition of bi- or \cb{trispectrum} Eqs. (\ref{eq:bispec_mode}) and 
(\ref{eq:tripsec_mode}) 
to reconstruct the generalized skew- and kurt-spectra.
This procedure gives a direct route to reconstruct the morphology of the CMB maps using modal decomposition.
The coefficients of modal decomposition are estimated using an orthogonal or separable basis function. This method
provides an alternative to the computation of MFs using generalized skew- and kurt-spectra of derivative maps
that we have developed in the text of the paper. This method of modal decomposition can work for generic multispectra
and thus can be useful to construct MFs or the related skew- or kurt-spectra in diverse cosmological situations.

{\bf 4. Skew-spectra for odd-parity Bispectrum:} 
We have also extended the concept of generalised skew-spectra to include the odd-parity
bispectrum in Eq. (\ref{eq:odd1})-Eq. (\ref{eq:odd3}). This will be useful in probing footprints of parity violating physics
in CMB maps. Even in the absence of any known parity-violating physics odd-parity 
skew-spectra can be useful for detecting systematic effects.

Finally we note that none of the derivations are based on any specific assumption about the nature of bispectrum.
Thus the generic results derived here will also be applicable to other
areas of cosmology where morphological estimators are used to estimate primordial
or secondary non-Gaussianity.
\section{Conclusion}
\label{sec:conclu}
We have generalized the concept of skew-spectra in different basis functions and used it to estimate the MFs. \ah{The aim 
is to define compressed non-Gaussianity statistics which retain information on the nature of the non-Gaussianity.}
This will allow cross-validation
of results obtained using different estimators at various intermediate steps, thus allowing a better
handle on any contamination from possible sources of systematics. The results we have derived are
independent of any specific assumptions regarding the nature of non-Gaussianity and can be useful
in other areas of cosmology. We have also included a contribution from odd-parity bispectrum 
in our reconstruction of MFs using the skew-spectra.

In addition to these statistics, we have developed an analytical framework which can be useful in 
estimating the S/N for a given experimental set up (beam and noise). Using this framework,
we found that, among the three skew-spectra probed, the S/N is
highest for $S_\ell^{(1)}$, with the ordering of $S_\ell^{(0)}$ and $S_\ell^{(2)}$ being model-dependent. 
We have tested four different models of non-Gaussianity. We find the estimators
$S_\ell^{(0)}$ and $S_\ell^{(1)}$ are highly anti-correlated beyond
$\ell = 30$ with a coefficient of correlation $r^{01}_{\ell}\approx-1$ for primordial non-Gaussianity. We found moderate correlation between
$S_\ell^{(0)}$ and $S_\ell^{(2)}$ as well as  $S_\ell^{(1)}$ and $S_\ell^{(2)}$ with $r_{02} \approx -0.5$ and 
$r_{12} \approx 0.5$ at $\ell\approx 1500$. We found the cumulative S/N on $f_{\rm NL}$ to be $\approx 0.1$
for the local model for $f_{\rm NL}=1$, scaling roughly proportionally to $f_{\rm NL}$.
The S/N can be improved by using
{\em Wiener-filtered} maps as inputs, and the results presented here can be generalized to take into
account such improvements.

The MFs provide a complementary tool to moment-based approaches in real space or equivalently 
multispectral analysis in the harmonic domain. Our results also provide an unifying approach
in different bases including in needlet basis.
Note that, our S/N results depend on various simplifying
approximations that allow analytical treatment; e.g. we have not included position-dependent
noise, which will involve hit-count maps, but recent studies have found the MFs to be rather insensitive to such detailed modelling \citep{Ducout12}.
We have adopted a $f_{\rm sky}$ approach for dealing with partial sky coverage;
detailed modelling will involve exact calculation of mode-mode coupling i.e. characterization
of galactic as well as a point source mask. In our derivation of scatter we have ignored all
higher-order correlations, which can always be characterized using numerical Monte Carlo simulations.
We do not include biases from secondaries which can generate spurious signatures independently or
through its coupling to primaries, e.g. generated by the ISW effect at large angular scales
or from the lensing and thermal Sunyaev-Zeldovich cross-correlation at smaller angular scales; both can provide
detectable observable signatures.

The skew-spectra that we have studied are mildly sub-optimal. However, we have developed 
generic reconstruction procedure for the MFs using optimal modal decomposition techniques that is 
typically used for construction of an optimum estimator. We also extend the method beyond the
bispectrum to take into account higher-order corrections to the level of trispectrum e.g. from lensing
of primary CMB.
\section{Acknowledgements}
\label{acknow}
DM acknowledges support
from STFC standard grant ST/G002231/1 at School of Physics and
Astronomy at Cardiff University where this work was completed. 
AC and JS are supported by NSF-AST0645427 and NASA NNX10AD42G.
DM would like to thank Michele Liguori for useful discussions.
We would like to thank an anonymous referee for many 
useful suggestions.
\bibliography{paper.bbl}
\appendix

\section{Minkowski Functionals and the CMB Sky}
\label{sec:mink_sky}
The discussion in the main text has been completely generic and is applicable to an arbitrary random 2D map on the surface of
the sky. We will specialize the discussion in this section to the case of CMB the
cleanest probes of primordial non-Gaussianity \citep{Planck13}, although the level of non-Gaussianity is
highly constrained by observation.

The angular multispectra for the temperature fluctuations
sample the 3D multispectra of the inflationary potential.
Given a specific form for the primordial non-Gaussianity,
it is possible to compute the MFs for the observed temperature perturbations.
The non-Gaussianity in the CMB sky can be a direct manifestation of the non-Gaussianity in the seed perturbations
generated during inflation. The non-Gaussianity in the inflationary potential is most easily characterized
in the Fourier domain, $\Phi({\bf k})$.
The following expression links the curvature fluctuations  $\Phi({\bf k})$ with 
spherical harmonic coefficients of the temperature anisotropy $a_{\ell m}$, with the help of the radiation transfer function  $\Delta_l(k)$ for
the temperature fluctuations \citep{WangKam00}. The angular power spectrum for the temperature fluctuations $\myC_\ell =\la a_{\ell m}a^*_{\ell m}\ra $ 
can be expressed in terms of the power spectrum of the 3D perturbations in the potential field 
$\la \Phi({\bf k}_1) \Phi({\bf k}_2)\ra = (2\pi)^3 \delta_{3D}({\bf k}_1 + {\bf k}_2)P_{\Phi}(k_1)$.
\be
a_{\ell m} = 4\pi (-i)^\ell \int {d^3k \over (2\pi)^3} \Phi(k)\Delta_\ell(k)Y_{\ell m}(\hat k); \quad
\myC_\ell = {2 \over \pi} \int k^2 dk P_\Phi(k) \Delta^2_{\ell}(k).
\ee
A Gaussian sky can be described statistically just by its angular power spectrum $\myC_\ell$. The lowest-order departure from the Gaussianity is described by the \cb{angular bispectrum}. The general form for the 3D bispectrum for the inflationary potential $\Phi$ is given as  
$B_{\Phi}(k_1,k_1,k_3) = (2\pi)^3 \delta_{3 \rm D}({\bf k}_1 + {\bf k}_2 + {\bf k}_3)$
$F_{\Phi}(k_1,k_2,k_3)$. In general translational invariance enforces momentum conservation in the Fourier domain leading
to the 3D Dirac delta function $\delta_{3D}$. The
kernel $F_{\Phi}(k_1,k_2,k_3)$ therefore is the amplitude of the bispectrum associated with each triangular configurations
involving the wave vectors ${\bf k}_i$. Various early Universe scenarios differ in $F_{\Phi}(k_1,k_2,k_3)$.
The reduced angular bispectrum defined in Eq. (\ref{eq:bispec_def}) can be expressed in terms of $F_{\Phi}(k_1,k_2,k_3)$:
\begin{equation}
b_{\ell_1\ell_2\ell_3} =  { \left ( 2 \over \pi \right ) }^3 \int dr r^2 
\int k_1^2 dk_1 j_{\ell_1}(k_1r)\Delta_{\ell_1}(k_1r)
\int k_2^2 dk_2 j_{\ell_2}(k_2r) \Delta_{\ell_2}(k_2r)
\int k_3^2 dk_3 j_{\ell_3}(k_3r)\Delta_{\ell_3}(k_3r) F_{\Phi}(k_1,k_2,k_3).
\end{equation}
Models of inflation can largely be divided into \cb{four} different categories. The first class of models is known as the local model
\citep{Salopek90,verdeheavens,  KomSpe01,MedeirosContaldi06,Crem03,Crem06,Cabella06,Liguori07,SmSeZa09}
and appears in multi-field models.
In these models the contribution to the bispectrum is maximum for the squeezed configurations i.e. when $ k_1 \ll k_2,k_3$.
The other main class of models are called equilateral models \citep{Chen06,Chen07}. In this class of models the maximum contribution
corresponds to a configuration where all wave vectors have similar magnitudes $k_1 \sim k_2 \sim k_3$ 
It is important to note that unlike the local model
the equilateral model can {\em not} represented by product of separable functions. However, approximate
separable forms do exist in the literature \citep{Crem06,SmZa06}. \cb{Notice that 
the local and equilateral forms are nearly orthogonal to each other and hence can be measured
nearly independently of each other. The other two models are known as {\em orthogonal}
and {\em enfolded} models respectively. The orthogonal model describes non-Gaussianity in single-field models 
with a non-canonical kinetic term and is nearly orthogonal to both the local and equilateral models.
The enfolded model is relevant for models with non Bunch-Davies vacuum
or general higher-derivative interactions.} We quote the results for the CMB non-Gaussianity
here, that arises in the context of local, equilateral, orthogonal and enfolded models. For more details
see e.g. \cite{KomSpe01,Komatsu_rev2010,YaWa10}.
\ben
&& b^{\rm loc}_{\ell_1\ell_2\ell_3} = 2 f_{\rm NL}\int r^2 dr [\beta_{\ell_1}(r)\beta_{\ell_2}(r)\alpha_{\ell_3}(r) + 
2\; {\rm cyc.perm.}]; \label{eq:model_loc}\\
&&b^{\rm equi}_{\ell_1\ell_2\ell_3} = 6 f_{\rm NL}^{\rm eq} \int r^2 dr [
-(\alpha_{\ell_1}(r)\beta_{\ell_2}(r)\beta_{\ell_3}(r) +2\; {\rm cyc.perm.}) - 2 \delta_{\ell_1}(r) \delta_{\ell_2}(r) \delta_{\ell_3}(r) 
+ (\beta_{\ell_1}(r) \gamma_{\ell_2}(r) \delta_{\ell_3}(r) + 5\;{\rm cyc.perm.})];\\
&& \cb{b^{\rm ortho}_{\ell_1\ell_2\ell_3} = 6 f_{\rm NL}^{\rm ortho} \int r^2 dr [
-3(\alpha_{\ell_1}(r)\beta_{\ell_2}(r)\beta_{\ell_3}(r) + 2\; {\rm cyc.perm.} ) - 8 \delta_{\ell_1}(r) \delta_{\ell_2}(r) \delta_{\ell_3}(r) + 
(3\beta_{\ell_1}(r)\gamma_{\ell_2}(r)\delta_{\ell_3}(r) + 5\; {\rm cyc.perm.})];} \\
&& \cb{b^{\rm en}_{\ell_1\ell_2\ell_3} = 6 f_{\rm NL}^{\rm en} \int r^2 dr [
(\alpha_{\ell_1}(r)\beta_{\ell_2}(r)\beta_{\ell_3}(r) + 2\; {\rm cyc.perm.} ) + 3 \delta_{\ell_1}(r) \delta_{\ell_2}(r) \delta_{\ell_3}(r) -
( \beta_{\ell_1}(r)\gamma_{\ell_2}(r)\delta_{\ell_3}(r) + 5\; {\rm cyc.perm.})].} 
\label{eq:model_eq}
\een
The following functions, used above, are useful in analytical expressions for the bispectrum and trispectrum \citep{Crem06}:
\ben
&& \cb{\alpha_\ell(r) = {2 \over \pi} \int  k^2 dk j_\ell(kr) \Delta_\ell(k)}; \quad
\beta_\ell(r) = {2 \over \pi} \int k^2 dk P_\Phi(k) j_\ell(kr) \Delta_{\ell}(k); \quad \\
&& \gamma_\ell(r)= {2 \over \pi} \int  k^2 dk P_\Phi^{1/3}(k) j_\ell(kr) \Delta_\ell(k); \quad
\delta_\ell(r)= {2 \over \pi} \int  k^2 dk P_\Phi^{2/3}(k) j_\ell(kr) \Delta_\ell(k); \quad \\
&& F_L(r_1,r_2) = {2 \over \pi} \int  k^2 dk P_\Phi^{2/3}(k) j_\ell(kr_1) j_\ell(kr_2).
\een
Here $j_\ell$ is a spherical Bessel function, 
and $\Delta_\ell(k)$
is the radiation transfer function which can be computed using the publicly available software such as CAMB\footnote{http://camb.info/} or 
CMBFAST\footnote{http://www.cmbfast.org/}. In addition to the bispectra the {\em reduced} CMB trispectrum in the local model
can be expressed in terms of these functions as
\ben
\tau^{\ell_1\ell_2}_{\ell_3\ell_4}(\ell)=&& 4f_{\rm NL}^2 h_{\ell_1\ell_2\ell}h_{\ell_3\ell_4\ell}\int r_1^2dr_1 \int r_2^2 dr_2 F_\ell(r_1,r_2) 
\alpha_{\ell_1}(r_1)\beta_{\ell_2}(r_1)\alpha_{\ell_3}(r_2)\beta_{\ell_4}(r_2)
\nonumber \\
&& + g_{\rm NL} h_{\ell_1\ell_2L}h_{\ell_3\ell_4L}\int r^2 dr \beta_{\ell_2}(r) \beta_{\ell_4}(r) 
[ \mu_{\ell_1}(r)\beta_{\ell_3}(r) + \mu_{\ell_3}(r)\beta_{\ell_1}(r)].
\label{eq:reduced}
\een
The above equation is derived from the following expression for the 3D trispectrum for the 
inflationary potential $\Phi$:
\ben
\cb{\la\Phi(k_1)\Phi(k_2)\Phi(k_3)\Phi(k_4)\ra} =&& \cb{(2\pi)^3 \delta_{3\rm D}({\bf k}_1 + {\bf k}_2 + {\bf k}_3 +{\bf k}_4)
\Big \{ {25 \over 9 } \tau_{\rm NL} \left [ P_{\Phi}(k_1)P_{\Phi}(k_2) P_{\Phi}(k_{13}) + 11 \; {\rm  cyc. perm.} 
\right ]}\nn \\
&& \cb{+ 6\; g_{\rm NL} \left [P_{\Phi}(k_1)P_{\Phi}(k_2) P_{\Phi}(k_3) + 3 \; {\rm cyc. perm.}  \right ]\Big\}; \quad\quad k_{ij} \equiv | {\bf k}_i+{\bf k}_j|}.
\label{eq:fnl_gnl}
\een
\cb{Assuming a curvature perturbation and standard local form one can derive $\tau_{\rm NL}=(6 f^{\rm loc}_{\rm NL}/5)^2$.
However, in a general inflationary scenario $\tau_{\rm NL}$ can be larger. Constraints on $\tau_{\rm NL}$
and $g_{\rm NL}$ were derived \citep{Smidt10,FRS10} using WMAP-5 data. Planck collaboration \citep{Planck13}
used maps from the nominal mission to constrain $\tau_{\rm NL}$ and found $\tau_{\rm NL}<2800$ ($95\%$ CL).}
For detailed discussions about various issues related to the symmetries and modelling of the CMB trispectrum see
\cite{huOka02,Hu00,Hu01,KomSpe01,Kogo06}.
%
%
\section{The Quadruplet of Kurt-Spectra and next to leading order corrections}
\label{sec:kurt}
\cb{The recent results from Planck satellite \citep{Planck13} indicate a low values of $f_{\rm NL}$ that characterize
different models of primordial bispectrum.
This motivates going beyond the lowest level of non-Gaussianity and using
the next-to-leading order trispectrum and the related quadruplet of kurt-spectra.
Indeed, several inflationary scenarios
exist in which the bispectrum is suppressed, and the trispectrum is the leading-order
non-Gaussianity in the data \citep{BSW06,Sasaki06,ByChoi10}. A detection of trispectra thus would be a very important validation of
such models. It is also important to realise that unlike one-point estimators,  the kurt-spectra
can separate out the amplitudes $\tau_{\rm NL}, g_{\rm NL}$ of two different types of topological diagrams, {\em snakes} and
{\em stars}, which contribute at the level of the trispectrum (see Eq. (\ref{eq:fnl_gnl}). This is interesting given recent that 
Planck results currently only constrain
$\tau_{\rm NL}$ and not $g_{\rm NL}$. From a different perspective, the kurt-spectra
can also be extremely useful in probing the lensing-induced topology changes that appear
at the level of the trispectrum.}

In a perturbative analysis, the leading-order terms that signify non-Gaussianity 
in the analysis of MFs depend on the bispectrum or equivalently a set of skewness terms.
The next-to leading-order correction terms depend on a set of {\em kurtosis} parameters $K^{(i)}$ that are fourth-order statistics 
and are analogues of the skewness parameters $S^{(i)}$ which we have defined above. In general the kurtosis parameters are collapsed fourth-order
one-point cumulants and probe the trispectrum with varying weights (see \cite{Mu_kurt10} for a more detailed discussion
on fourth-order one-point cumulants, their two-point counterparts, the cumulant correlators,  and the related harmonic-space statistics). 
The four different kurtosis parameters 
that are related to the MFs are a natural generalisation of the ordinary kurtosis $K^{(0)}$ which
is routinely applied in many cosmological studies. We will denote these generalised kurtosis parameters by $K^{(i)}; i = 1,2,3$.
These parameters are constructed from the derivative field of the original map
map $\Psi(\oh)$ and its derivatives  $\nabla\Psi(\oh)$ and $\nabla^2\Psi(\oh)$ as follows.  
\ben
&& K^{(0)} \equiv {1 \over \sigma_0^6}K^{(\myf^4)} = {\la \myf^4 \ra_c \over \sigma_0^6}; \quad
K^{(1)} \equiv {1 \over \sigma_0^4 \sigma_1^2 } K^{(\myf^3\nabla^2 \myf)} = {\la \myf^3 \nabla^2 \myf \ra_c \over \sigma_0^4 \sigma_1^2}; \quad \\
&& K^{(2)} \equiv  K^{(2a)} +   K^{(2b)} \equiv {{\cb 2} \over \sigma_0^2\sigma_1^4} K^{(\myf |\nabla \myf |^2 \nabla^2 \myf)} + {1 \over \sigma_0^2\sigma_1^4} K^{(|\nabla\myf|^4)}  =
\cb{2} {\la \myf |(\nabla \myf)|^2 (\nabla^2\myf) \ra_c  \over \sigma_0^2\sigma_1^4}
+ {}{\la |(\nabla \myf)|^4 \ra_c  \over \sigma_0^2 \sigma_1^4};\\
&& \quad K^{(3)} \equiv {1 \over 2\sigma_0^2\sigma_1^4} K^{(|\nabla\myf|^4)} = {\la |\nabla\myf|^4 \ra_c \over 2\sigma_0^2\sigma_1^4}.
\label{kurtosis_real_space}
\een
The subscript $_c$ \cb{corresponds} to the connected components; Gaussian contributions
are subtracted out, including both noise and signal. The evaluation of these moments is 
relatively easy in real space for a pixelised map, and involves taking derivatives of beam-smoothed maps.
The corresponding power spectra associated with these fourth-order moments are constructed by 
cross-correlating appropriate maps in the harmonic domain and are easy to implement numerically (Eq. (\ref{eq:kurt_spectra}) provides 
exact expressions for the corresponding power spectra or the kurt-spectra). 

The next-to-leading order corrections to the MFs involve these $K^{(i)}$s as well as the product of two
skewness parameters $S^{(i)}$ (i.e. the terms such as $[S^{(0)}]^2$, $[S^{(0)}S^{(1)}]$ or $[S^{(1)}S^{(2)}]$)
defined previously in the context of leading order non-Gaussian terms \citep{Mat10}.  
\cb{The next-to-leading order corrections $v_k^{(3)}(\nu)$ introduced in Eq.(\ref{eq:small_vs}) can be expressed as follows:} 
\ben  
&& {v}_0^{(3)}(\nu) = {[S^{(0)}]^2 \over 72}{\cal H}_5(\nu) +{K^{(0)} \over 24} {\cal H}_3(\nu); \quad \label{eq:fourth_mink_0}\\
&& {v}_1^{(3)}(\nu) = {[S^{(0)}]^2 \over 72}{\cal H}_6(\nu) +\left [ {K^{{(0)}}-S^{(0)}S^{(1)} \over 24} \right ] {\cal H}_4(\nu) -{1 \over 12}\left [ K^{\cb{(1)}} + {3 \over 8} 
[S^{(1)}]^2 \right ] {\cal H}_2(\nu) - {1 \over 8} K^{(3)} \\
&& {v}_2^{(3)}(\nu) = {[S^{(0)}]^2 \over 72}{\cal H}_7(\nu) + \left [ {K^{(0)} - \cb{2} S^{(0)}S^{(1)} \over 24} \right ] 
{\cal H}_5(\nu) -{1 \over 6}\left[ K^{(1)} + {1 \over 2}S^{(0)}S^{(2)} \right] {\cal H}_3(\nu) - {1 \over 2}\left [ K^{(2)} + {1 \over 2} S^{(1)}S^{(2)} \right ]{\cal H}_1(\nu).
\label{eq:fourth_mink}
\een
The analytical modelling of four-point correlation functions is most naturally done in the harmonic domain.
They are described by the angular trispectrum $T^{\ell_1\ell_2}_{\ell_3\ell_4}(\ell)$, which is defined through the relation
\cb{
\be
\la a_{\ell_1m_1}a_{\ell_2m_2}a_{\ell_3m_3}a_{\ell_4m_4}\ra_c = \sum_{\ell m}(-)^M\left (\begin{array}{ c c c }
     \ell_1 & \ell_2 & \ell\\
     m_1 & m_2 & m
  \end{array} \right)\left ( \begin{array}{ c c c }
     \ell_3 & \ell_4 & \ell\\
     m_3 & m_4 & -m
  \end{array} \right) T^{\ell_1\ell_2}_{\ell_3\ell_4}(\ell).
  \ee}
The trispectrum $T^{\ell_1\ell_2}_{\ell_3\ell_4}(\ell)$ is expressed in terms of the \cb{pairing function} $P^{\ell_1\ell_2}_{\ell_3\ell_4}(\ell)$, 
encoding all possible inherent symmetries \citep{Hu01}.                      
\be 
T^{\ell_1\ell_2}_{\ell_3\ell_4}(\ell) = P^{\ell_1\ell_2}_{\ell_3\ell_4}(\ell) 
+ \Xi_{\ell}\left [\sum_{\ell'} (-1)^{\ell_2+\ell_3}\left \{ \begin{array}{ c c c }
     \ell_1 & \ell_2 & \ell \\
     \ell_4 & \ell_3 & \ell'
  \end{array} \right \} P^{\ell_1\ell_3}_{\ell_2\ell_4}(\ell') 
+ \sum_{\ell'}  
(-1)^{\ell+\ell'} \left \{ \begin{array}{ c c c }
     \ell_1 & \ell_2 & \ell \\
     \ell_3 & \ell_4 & \ell'
  \end{array} \right \} P^{\ell_1\ell_4}_{\ell_3\ell_2}(\ell') \right ].
\label{Total_Tri}
\ee
\n The matrices in curly brackets are $6j$ symbols which are defined using $3j$ symbols 
(see \cite{Ed68} for more detailed discussions). The entities $P^{\ell_1\ell_2}_{\ell_3\ell_4}(\ell)$ can be further decomposed in terms of  
the {\it reduced} function $\tau^{\ell_1\ell_2}_{\ell_3\ell_4}(\ell)$:
\be
 P^{\ell_1\ell_2}_{\ell_3\ell_4}(\ell) = \tau^{\ell_1\ell_2}_{\ell_3\ell_4}(\ell)
+ (-1)^{\Sigma_U}\tau^{\ell_2\ell_1}_{\ell_3\ell_4}(\ell) + (-1)^{\Sigma_L} \tau^{\ell_1\ell_2}_{\ell_4\ell_3}(\ell) + 
(-1)^{\Sigma_L + \Sigma_U}\tau^{\ell_2\ell_1}_{\ell_4\ell_3}(\ell); \quad \Sigma_{\cb U} = \ell_1+\ell_2+\ell; \quad \Sigma_{\cb{L}} = \ell_3+\ell_4+\ell.
\ee
Each individual model for primordial non-Gaussianity makes a specific prediction for the {\it reduced} trispectrum
which can be used as a fingerprint to rule out many possibilities.

Next we will introduce three additional trispectra that are constructed using different weights to the original trispectra 
$T^{\ell_1\ell_2}_{\ell_3\ell_4}({\cb{\ell}})$ and differ in the way they weight various modes,that are specified by a particular choice of the quadruplet $\{ \ell_i \}$. 
\ben
&& [T^{(0)}]^{\ell_1\ell_2}_{\ell_3\ell_4}(\ell) =  T^{\ell_1\ell_2}_{\ell_3\ell_4}(\ell); \quad
 [T^{(1)}]^{\ell_1\ell_2}_{\ell_3\ell_4}(\ell) = {1 \over 4} \left [{{\Pi}_{\ell_1}+{\Pi}_{\ell_2}+{\Pi}_{\ell_3}+{\Pi}_{\ell_4}}\right ] 
T^{\ell_1\ell_2}_{\ell_3\ell_4}(\ell); \label{eq:kurt_first} \\
&& [T^{(2)}]^{\ell_1\ell_2}_{\ell_3\ell_4}(\ell) = {1 \over 4}\left [{{\Pi}_{\ell}- ({\Pi}_{\ell_1}+{\Pi}_{\ell_2})({\Pi}_{\ell_3}+{\Pi}_{\ell_4}) }
\right ] T^{\ell_1\ell_2}_{\ell_3\ell_4}(\ell); \\
&& [T^{(3)}]^{\ell_1\ell_2}_{\ell_3\ell_4}(\ell) = {1 \over 4}\left [ {({\Pi}_{\ell_1}+ {\Pi}_{\ell_2}-{\Pi}_{\ell})({\Pi}_{\ell_3}+{\Pi}_{\ell_4}-{\Pi}_{\ell})}
\right ] T^{\ell_1\ell_2}_{\ell_3\ell_4}(\cb{\ell}).
\label{eq:kurt_spectra1}
\een
The four generalised kurtosis and the related kurt-spectra we have defined above can now be expressed in terms of these generalised trispectra $T^{(i)}$ as follows:
\be
K^{(i)} = \sum_{\ell_i}\sum_{\ell} [T^{(i)}]^{\ell_1\ell_2}_{\ell_3\ell_4}(\ell) I_{\ell_1\ell_2\ell} I_{\ell_3\ell_4\ell}; \quad
K^{(i)}_\ell = \sum_{\ell_i}{1 \over \Xi_\ell}  [T^{(i)}]^{\ell_1\ell_2}_{\ell_3\ell_4}(\ell) J_{\ell_1\ell_2\ell} J_{\ell_3\ell_4\ell}; \quad
K^{(i)} = \sum_{\ell}\; \Xi_{\ell} \; K^{(i)}_\ell.
\label{eq:kurt_spectra2}
\ee
These kurt-spectra can be estimated using techniques that are very similar to
techniques we have employed to estimate the skew-spectra before; e.g. to construct the first of
these kurt-spectra we have to cross-correlate the squared field $\Psi^2$ with it itself. Other 
kurt-spectra are similarly constructed by cross-correlating quadratic constructs which also involve the
derivative maps: 
\ben
&& K^{(0)}_\ell \equiv {1 \over \Xi_{\ell}}\sum_m  [\Psi^2]_{\ell m}[\Psi^2]^*_{lm}; \quad K^{(1)}_\ell \equiv {1 \over \Xi_{\ell}}\sum_m  [\Psi^2]_{\ell m}[\Psi\nabla^2\Psi]^*_{\ell m}; \label{eq:direct1}\\
&& K^{(2a)}_\ell \equiv {1 \over \Xi_{\ell}}\sum_m  [\Psi^2]_{\ell m}[\nabla^2\Psi]^*_{\ell m}; \quad 
K^{(2b)}_\ell \equiv {1 \over \Xi_{\ell}}\sum_m  [\Psi^2]_{lm}[\Psi\nabla^2\Psi]^*_{\ell m}; \quad K^{(3)}_\ell \equiv {1 \over \Xi_{\ell}}\sum_m [|\nabla\Psi|^2]_{\ell m}[|\nabla\Psi|^2]^*_{\ell m}.
\label{eq:kurt_spectra}
\een
\n
We have defined two different estimators  $K^{(2a)}_\ell$ and  $K^{(2b)}_\ell$ that can be jointly used to construct the kurt-spectra 
$K^{(2)}_\ell$. The treatment for the masked sky follows exactly the same manner.  For individual  $K^{(i)}_\ell$ the
unbiased estimators can be recovered using exactly the same mode-coupling matrix $M_{\ell\ell'}$ introduced before in 
Eq. (\ref{eq:auto_cov}); we have for the
masked kurt-spectra $\hat K^{(i)}_\ell = {\mathbb M}^{-1}_{\ell\ell'} \tilde K^{(i)}_{\ell'}$. The auto- and the covariance of these estimators
can also be estimated using obvious generalization of Eq. (\ref{eq:auto_cov}) and Eq. (\ref{eq:cross_cov}) respectively.

The fourth-order expressions for the power spectra associated with MFs \cb{$[v^{(3)}_k]_\ell(\nu)$} can be obtained by 
replacing all one-point $K^{(i)}$s with their two-point counterparts $K^{(i)}_\ell$ in Eq. (\ref{eq:fourth_mink_0})-Eq. (\ref{eq:fourth_mink}) 
\cb{(see Eq. (\ref{eq:nu_spec}) for a similar expression for \cb{$[v^{(2)}_k]_\ell(\nu)$} 
in terms of the generalised skewness parameters)}.
The contributions from the skewness parameters only contribute in the monopole terms. The  extraction of the kurt-spectra in the presence of a
mask can be carried out analogously to the skew-spectra.

Corresponding cumulant correlators in the real space are given by: $\myC^{\Psi^2,\Psi^2}(\theta_{12})$, $C^{\Psi^2,\Psi\nabla^2\Psi}(\theta_{12})$,
$\myC^{\nabla\Psi\cdot\nabla\Psi,\Psi\nabla^2\Psi}(\theta_{12})$ and  $\myC^{\nabla\Psi\cdot\nabla\Psi,\nabla\Psi\cdot\nabla\Psi}(\theta_{12})$.
In the limit of zero separation they collapse to their one-point counterpart, i.e. $\myC^{\Psi^2,\Psi^2}(0)=K^{(0)}$, 
$\myC^{\nabla\Psi\cdot\nabla\Psi,\Psi\nabla^2\Psi}(0)=K^{(1)}$, $\myC^{\nabla\Psi\cdot\nabla\Psi,\nabla\Psi\cdot\nabla\Psi}(0)=K^{(2)}$
and $\myC^{\nabla\Psi\cdot\nabla\Psi,\nabla\Psi\cdot\nabla\Psi}(0)= K^{(3)}$.

On a different note, the power spectra associated with the kurtosis or kurt-spectra are discussed in detail for
a scalar field \citep{Mu_kurt10}, where two different types of kurt-spectra were introduced
in the context of analysis of CMB Temperature maps. These two kurt-spectra $K_\ell^{(2,2)}$ and $K_\ell^{(3,1)}$ 
both sample the relevant trispectrum. The first of these is constructed from cross-correlating the squared map with itself. 
The other kurt-spectrum is constructed from cross-correlating a cubic map with the original map. 
In general different sets of maps can also be used to form squared and cubic combinations which will 
probe a mixed trispectra. In the present context we are interested in the spectra  $K_\ell^{(2,2)}$, as the construction of 
$K_\ell^{(3,1)}$ will involve gradient maps and are more complicated to analyse in a coordinate-independent way. 
However, such constructions are indeed possible using the spinorial formalism.
\begin{figure}
\begin{center}
{\epsfxsize=15. cm \epsfysize=7.25cm {\epsfbox[35 326 578 652]{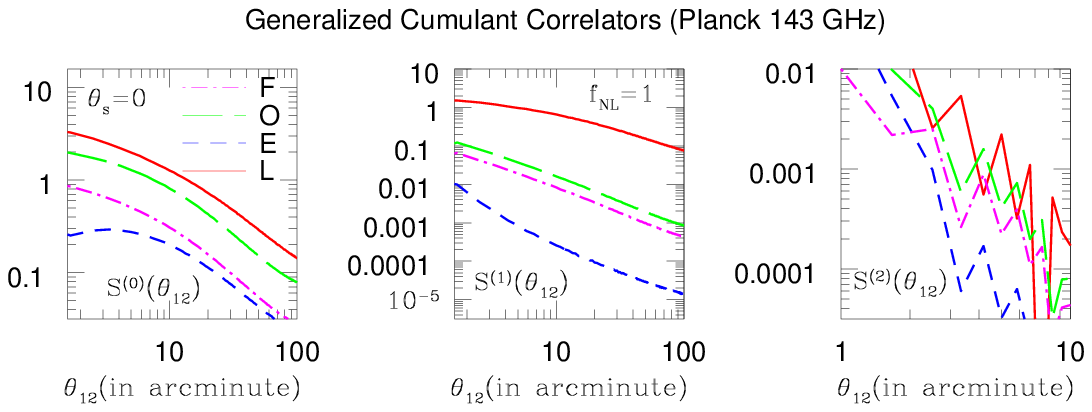}}}
\caption{The cumulant correlators defined in Eq. (\ref{eq:cumu1})-Eq. (\ref{eq:cumu3}) are plotted as a function
of the separation angle $\theta_{12}$ (in arcminute) for the various models of non-Gaussianity for 
Planck (143 GHz channel) experiment as indicated. The cumulant correlators 
and the corresponding skew-spectra carry equivalent information. Various models of primordial non-Gaussianity
considered are {\em local} (L), {\em equilateral} (E), {\em enfolded} (F) and {\em orthogonal} (O).}
\label{fig:cumu_loc}
\end{center}
\end{figure} 
The physical meaning of these kurt-spectra can be understood more easily 
in the harmonic domain. As mentioned, each individual mode of the trispectrum is characterized by 
a specific choice of the set of modes ${\ell_i}$ that defines it. These modes each constitute the sides of a quadrilateral 
whose diagonal is specified by the quantum number $\ell$.  Note that, the kurt-spectra that we have considered 
here take contributions from all possible configurations of the quadrilateral while keeping
its diagonal $\ell$ fixed. 

The estimation of the kurt-spectra from real data is relatively easy, and follows the same
methodology as the skew-spectra. The first of these kurt-spectra $K_{\ell}^{(0)}$ is extracted by cross-correlating
the squared field $\myf^2(\oh)$ with itself. The spectrum  $K_{\ell}^{(1)}$ is constructed by
cross-correlating $\myf^2(\oh)$ with $\myf\nabla^2\myf$. The other two kurt-spectra can
likewise be constructed. In each such construction a scalar map from a product field is 
generated before it is cross-correlated with another such map. 

The corrections to the power spectrum associated with the MFs now can be written in terms of the
$K^{(i)}_\ell$, i.e. the kurt-spectra and the various Hermite polynomials as introduced above.
The contributions from the lower-order statistics such as skewness will only
contribute to the monopole term for every MFs. However, the higher \cb{multipoles} will involve
contributions from various kurt-spectra as indicated in Eq. (\ref{eq:fourth_mink_0})-Eq. (\ref{eq:fourth_mink}). Different specific choice of trispectra
will therefore lead us to completely different power spectra associated with the MFs
and can help to distinguish various models of non-Gaussianity.

The generalized skew- and kurt-spectra can also be useful in probing the detection of topological defects
through their effect on change in topology of the CMB temperature \citep{RS10}.
\section{3j Symbols}
\label{3j}
We list here various expressions related to $3j$ symbols \citep{Ed68} that were used in the text.
\ben
&& \left ( \begin{array}{ c c c }
     \ell_2 & \ell_1 & \ell_3 \\
     m_2 & m_1 & m_3
  \end{array} \right) =
(-1)^{\ell_1+\ell_2+\ell_3}\left ( \begin{array}{ c c c }
     \ell_1 & \ell_2 & \ell_3 \\
     m_1 & m_2 & m_3
  \end{array} \right)
\label{Ortho0}\\
&& \sum_{\ell_3m_3} (2\ell_3+1) \left ( \begin{array}{ c c c }
     \ell_1 & \ell_2 & \ell_3 \\
     m_1 & m_2 & m_3
  \end{array} \right )
\left ( \begin{array}{ c c c }
     \ell_1 & \ell_2 & \ell \\
     m_1' & m_2' & m
  \end{array} \right ) = \delta_{m_1m_1'} \delta_{m_2m_2'} 
\label{Ortho1}\\
&& \sum_{m_1m_2} \left ( \begin{array}{ c c c }
     \ell_1 & \ell_2 & \ell_3 \\
     m_1 & m_2 & m_3
  \end{array} \right)
\left ( \begin{array}{ c c c }
     \ell_1 & \ell_2 & \ell_3' \\
     m_1 & m_2 & m_3'
  \end{array} \right) = {\mathcal \delta_{\ell_3l_3'} \delta_{m_3m_3'} \over 2\ell_3 + 1} 
\label{Ortho2}\\
&& (-1)^m\left ( \begin{array}{ c c c }
     \ell & \ell & \ell' \\
     m & -m & 0
  \end{array} \right ) = {(-1)^\ell \over \sqrt{(2\ell+1)}} \delta_{\ell'0}
\label{special1}
\een
\n
The Gaunt (or overlap) integral involving three spherical harmonics can be expressed in terms of $3j$ symbols:
\be
\int d\oh~ Y_{\ell_1m_1}(\oh) Y_{\ell_2m_2}(\oh) Y_{\ell_3m_3}(\oh) = \sqrt{(2\ell_1+1)(2\ell_2+1)(2\ell_3+1)\over 4 \pi}\left( \begin{array}{ c c c }
     \ell_1 & \ell_2 & \ell_3 \\
     0 & 0 & 0
  \end{array} \right )
\left ( \begin{array}{ c c c }
     \ell_1 & \ell_2 & \ell_3 \\
     m_1 & m_2 & m_3
  \end{array} \right )
\label{eq:gaunt}
\ee
\n
\end{document}